\documentclass{article}
\usepackage{float}
\usepackage{subcaption}
\usepackage{fullpage}

\usepackage{setspace}
\usepackage{stfloats}
\usepackage[normalem]{ulem}
\usepackage{algorithm}
\usepackage{algpseudocode}

\usepackage{indentfirst}
\usepackage{parskip}
\usepackage{titlesec}
\usepackage[section]{placeins}

\usepackage{breakcites}
\usepackage{hyphenat}
\usepackage[utf8]{inputenc}  
\usepackage[T1]{fontenc}     
\usepackage{csquotes}  
\usepackage{xcolor}

\usepackage[a4paper,margin=1.31cm, marginparwidth=0cm]{geometry}

\usepackage[backend=biber]{biblatex}

\PassOptionsToPackage{hyphens}{url}
\usepackage[colorlinks = true,
            linkcolor = blue,
            urlcolor  = blue,
            citecolor = blue,
            anchorcolor = blue]{hyperref}
\usepackage{etoolbox}

\renewenvironment{abstract}
  {{\bfseries\noindent{\abstractname}\par\nobreak}\footnotesize}
  {\bigskip}

\titlespacing{\section}{0pt}{*3}{*1}
\titlespacing{\subsection}{0pt}{*2}{*0.5}
\titlespacing{\subsubsection}{0pt}{*1.5}{0pt}

\usepackage{authblk}

\usepackage{graphicx}
\usepackage[space]{grffile}
\usepackage{latexsym}
\usepackage{textcomp}
\usepackage{longtable}
\usepackage{tabulary}
\usepackage{booktabs,array,multirow}
\usepackage{amsfonts,amsmath,amssymb}
\providecommand\citet{\cite}
\providecommand\citep{\cite}

\newif\iflatexml\latexmlfalse

\AtBeginDocument{\DeclareGraphicsExtensions{.pdf,.PDF,.eps,.EPS,.png,.PNG,.tif,.TIF,.jpg,.JPG,.jpeg,.JPEG}}

\usepackage[utf8]{inputenc}
\usepackage[english]{babel}

\begin{document}

\title{Reconstructing MSM Sexual Networks to Guide PrEP Distribution Strategies for HIV Prevention}

\author[1,2,*]{João Brázia}%
\author[4,5]{Istv\'an Z. Kiss}%
\author[6]{Alexandre P. Francisco}%
\author[1,2,3]{Andreia Sofia Teixeira}%
\affil[1]{BRAN Lab, Network Science Institute, Northeastern University London, London, UK}%
\affil[2]{Kent Medway Medical School, Canterbury, UK}%
\affil[3]{LASIGE, Faculdade de Ciências, Universidade de Lisboa, Portugal}%
\affil[4] {Network Science Institute, Northeastern University London, London, UK}
\affil[5]{Department of Mathematics, Northeastern University, Boston, USA}
\affil[6]{INESC-ID, Instituto Superior Técnico, Universidade de Lisboa, Portugal}

\vspace{-1em}

\begingroup
\let\center\flushleft
\let\endcenter\endflushleft
\maketitle
\endgroup

\selectlanguage{english}
\begin{abstract}
{Men who have sex with men (MSM) remain disproportionately affected by HIV, yet optimizing Pre-exposure Prophylaxis (PrEP) distribution remains a public health challenge. Current guidelines and most modelling studies do not incorporate sociodemographic or network-level factors that shape transmission. While network reconstruction from egocentric data has been studied, the relative importance of demographic mixing dimensions remains uncertain. Using data from 4,667 MSM participants, we show that uncertainty in network reconstruction from egocentric survey data — specifically whether assortativity by age or race is incorporated — affects simulated HIV prevalence under the same observed PrEP uptake. We simulate HIV transmission over 50 years across this structural space and evaluate whether empirically observed uptake reaches transmission-critical network positions. Network structure strongly influences outcomes: assortative by degree networks show 17\% lower equilibrium prevalence due to hub isolation within communities. Targeted PrEP strategies based on degree or k-shell centrality achieved the highest prevalence reductions, particularly in assortative by age and race networks where hubs bridge demographic groups. PrEP uptake from data is suboptimal in assortative by age and race networks, underperforming compared with network-based strategies. Results demonstrate that uncertainty in network reconstruction affects intervention design and highlight the need for robust prevention strategies under structural ambiguity.

}%
\end{abstract}%


\setlength{\parindent}{10pt}
\setlength{\parskip}{0pt}

\section*{Introduction}
Men who have sex with men (MSM) are at a disproportionate risk of HIV infection, accounting for 67\% of sexually transmitted HIV cases in the United States \cite{world2023hiv}. This burden reflects not only biological susceptibility associated with the nature of sexual intercourse but also complex behavioral and structural factors, including partner concurrency, condomless sex, substance use, and elevated HIV prevalence. While individual and behavioral factors contribute to HIV risk, effective prevention must also consider the network-level structures that sustain transmission.

Pre-exposure prophylaxis (PrEP) has demonstrated high efficacy for HIV prevention in MSM across multiple trials and population studies \cite{grant2010preexposure, mayer2020emtricitabine, fonner2016effectiveness}. Nevertheless, PrEP’s efficiency is highly dependent on the choice of the most appropriate eligibility guidelines. These depend on the estimation of the risk of exposure to HIV based solely on sex-related behaviors and vary between countries \citep{wang2025comparing}. For example, current U.S. guidelines prioritize PrEP for individuals who, in the last six months, have reported sexual intercourse with HIV-positive partners, other sexually transmitted diseases (STDs), and/or inconsistent condom use\cite{cdc2021prep}. However, individual-level risk behaviors may not capture contact patterns and network-level transmission risk. An individual with moderate behavioral risk but high network centrality, bridging multiple communities or connecting to many partners, may pose greater transmission risk than someone with high-risk behaviors but few connections. Because sexual network structure is not directly observable in routine surveillance, PrEP allocation is designed without knowledge of individuals’ positions in the contact network, leaving open the question of whether current uptake reaches the most transmission-critical individuals~\cite{vissers2008impact, gantenberg2018improving, kasaie2017impact, pastor2002immunization}. Nonetheless, previous work has shown that risk-based targeting is not always the most effective in practice, particularly in heterogeneous efficacy scenarios \cite{steinegger2022non}.

Network-based models studies have provided valuable insights on optimized strategies to allocate PrEP among MSM, suggesting that targeting highly connected individuals can 
improve intervention efficiency \cite{bernini2019evaluating, choi2020cost, nichols2016cost, anderson1989mathematical, punyacharoensin2016effect}. However, existing approaches face critical limitations: they either require complete contact data, which is extremely difficult to collect for stigmatized populations, or rely on simplified topologies that fail to capture real-world network features. Moreover, MSM sexual networks exhibit distinctive structural properties that fundamentally influence transmission but remain absent from intervention planning frameworks. In particular, studies have found that people tend to connect or choose potential sexual partners within the same age or racial groups \cite{abuelezam2019interaction, amirkhanian2014social}. Members of the same sexual network often exhibit similar norms, attitudes, and HIV risk behavior levels \cite{amirkhanian2014social}. Network-tailored public health policies could substantially reduce transmission chains among high-risk individuals by identifying and prioritizing those whose position in a given network structure, along with individual behavior, makes them critical to epidemic dynamics. 

\begin{figure*}[!hbt]
\centering
\includegraphics[width=1.0\textwidth]{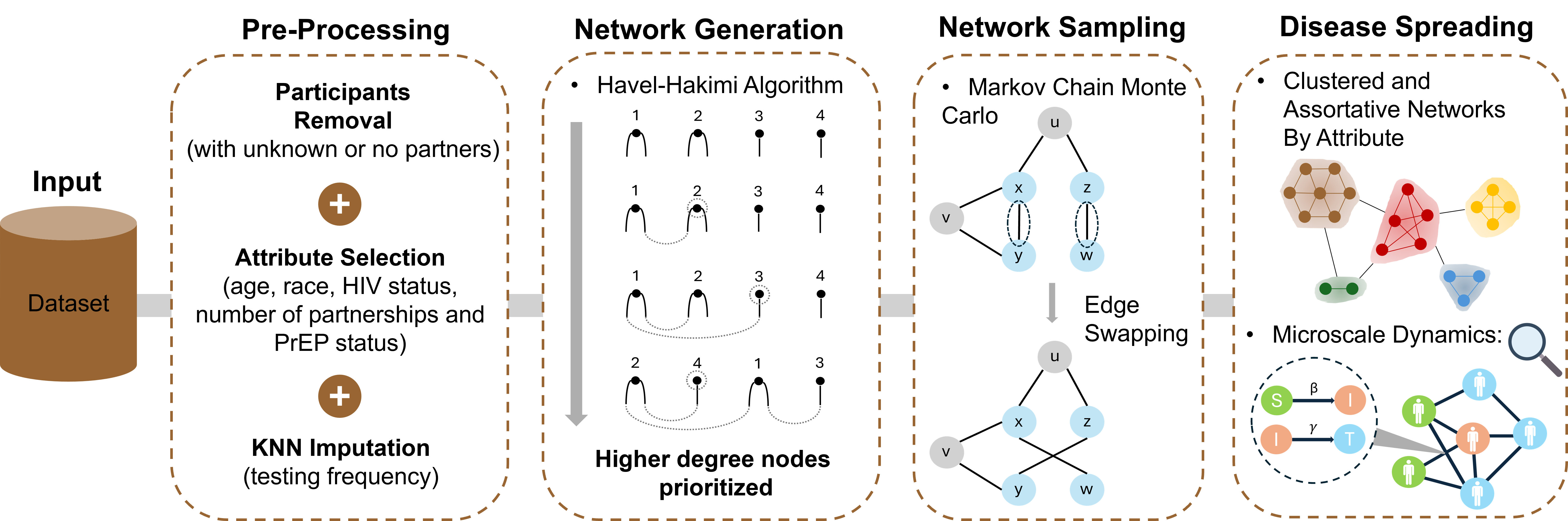}
\caption{\textbf{Workflow from data preprocessing to simulations of epidemic spreading on networks and deployment of PrEP under different intervention scenarios.}}
\label{fig:workflow_network_generation}
\end{figure*}

In this work, we leverage data from 4,667 MSM participants in the ARTnet study~\cite{weiss2020egocentric}, of whom 13.67\% were on PrEP at the time of data collection, to: (i) develop and validate degree-preserving network reconstruction algorithms that uniquely incorporate demographic assortativity by age, race, and sexual activity, thus leading to networks with realistic structures; (ii) simulate HIV transmission dynamics across networks with different topological properties; (iii) compare PrEP allocation strategies based on network centrality — degree-based and k-shell decomposition — against random distribution across this structural space, quantifying how network topology modulates intervention effectiveness; and (iv) assess whether observed PrEP uptake from the data takes transmission-critical network positions, or whether there is an under-representation of individuals occupying central or bridging roles across different network reconstructions (see Figure~\ref{fig:workflow_network_generation}).

Our findings show that networks reconstructed assuming assortative mixing with respect to different attributes, including age, race and activity level, can differ significantly, leading to different HIV transmission patterns and PrEP effectiveness, with important implications for public health policy. Targeted strategies, focusing on centrality measures, consistently outperform observed PrEP uptake from the data, particularly in networks where highly connected individuals bridge demographic groups. These results provide evidence of the need for a practical framework when complete network data is unavailable, highlighting opportunities to suggest improvements in HIV prevention and optimized PrEP distribution in MSM communities.

\section*{Results}

\subsection*{Data Overview}
We use data from the ARTnet study \cite{weiss2020egocentric}, an anonymous cross-sectional Internet-based survey that collected information on sociodemographic characteristics, HIV-related risk behavior, testing, and the use of prevention strategies or services of 4904 MSM in the United States, between 2017 and 2019. To take part in the study, participants had to identify as a male cisgender and biologically male, have a history of sexual intercourse with other men, and be between 15 and 65 years old. After preprocessing the data (see Methods Section A), we obtained a cohort of 4667 MSM from which we extracted age, HIV status, number of sexual partnerships, race, PrEP status, and testing frequency \cite{weiss2020egocentric}.

With this cohort, we create a synthetic population that expresses real demographic and behavioral characteristics sampled from a MSM community in the United States. The median (P25-P75) number of partnerships and the testing frequency are, respectively, 4 (2-10) and 2 (1-4). The most prevalent race group is white (72\%), followed by hispanic (14\%), other (9\%), and black (5\%). The median (P25-P75) age is 32 (24-49). The proportion of participants reporting being currently on PrEP was 13.67\%, consistent with 2017 CDC estimates for MSM \cite{CDCPrEPFactSheet}. Complete distributions of number of contacts, age, race, and testing frequency are given in the Supplementary Information Fig. S3.

\subsection*{Network reconstruction from sociodemographic data}
The ARTNet dataset contains only individual-level information without contact patterns structure. To overcome this limitation, we propose a reconstruction framework that generates synthetic sexual networks from sociodemographic data and explores how network structure changes when assortativity is imposed with respect to different demographic attributes individually. First, we implemented the Havel-Hakimi algorithm \cite{havel1955remark,hakimi1962realizability} to generate networks that preserve the observed degree sequence ($V=4,667$ nodes, $E=25,107$ edges), followed by a Markov Chain Monte Carlo (MCMC)-based edge swapping to ensure proper randomization (see Methods Section B, Supplementary Information Algorithm 1 and 2). To ensure statistical robustness, we run 10 independent Markov chains, each initialized with 10 different networks generated by the Havel-Hakimi algorithm. Each chain runs until convergence is reached. This ensemble of 100 random networks serves as a baseline for independently modifying specific structural properties while keeping the degree distribution fixed. Real sexual contact networks reflect not only who connects to whom (how many partners -- degree), but most importantly how individuals tend to connect to those who share similar characteristics.\cite{amirkhanian2014social}. 
Thus, beyond common approaches that generate networks with clustering and degree assortativity~\cite{bansal2009exploring,newman:assortativity_mixing,van2010influence}, we incorporate assortative mixing by demographic attributes, specifically age and race, into network reconstruction, revealing their distinct structural effects on HIV transmission dynamics. By reconstructing networks based on these four network properties, we create networks that range from random mixing to highly segregated demographic communities (see Methods Section B and Supplementary Information Section 4).

\subsection*{Network structure: from a global to a mesoscopic perspective}

To examine how assortativity by different attributes shapes network structure beyond degree distributions, we compare networks reconstructed using traditional methods (clustering and degree assortativity) with our demographic-based approach (age and race assortativity). We characterize such networks across multiple scales, from global topology to mesoscale community organization. 
\par At the global-scale level, the network topology of each generated network is characterized by computing the largest component (LC), average path length ($\langle l \rangle$), diameter ($D$), and modularity ($Q$) while systematically varying the structural property of interest (see Supplementary Information Fig. S5 and S6 (a)). Modularity was estimated by partitioning the network using the Louvain algorithm, as implemented in the \textit{python-louvain} package \cite{python_louvain}. All remaining global metrics were computed using \textit{NetworkX}. Table~\ref{tab:final_metrics} shows the metrics of networks with maximal clustering and maximal assortativity by degree, age, and race, averaged across all network realizations. We observe that assortative by degree and clustered networks exhibit the most pronounced structural changes, with the longest average path lengths ($ \langle l \rangle = 6.23 \pm 0.0526$ and $4.84 \pm 0.0192$, respectively) and bigger diameters ($D = 21.6 \pm 1.84$ and $12.3 \pm 0.853$), while maintaining substantial connectivity despite increased fragmentation. Modularity remains similar across clustered ($Q=0.558 \pm 0.00435$), assortative by degree ($Q=0.573 \pm 0.00902$), and assortative by age ($Q=0.601 \pm 0.00782$) networks. In contrast, assortative by race networks show minimal structural deviation from the random networks, presenting the lowest fragmentation ($DC = 11.6 \pm 3.33$) and modularity ($Q=0.373 \pm 0.00353$), suggesting weaker racial homophily in sexual contact patterns.

\begin{table*}[t]
\caption{\textbf{Structural metrics of networks with maximally enhanced properties.} From an ensemble of 100 baseline networks (generated via 10 MCMC chains × 10 initial Havel–Hakimi network configurations), we optimized each property through edge rewiring and selected those networks attaining the largest increase in the target metric. Values show the mean and standard deviation of the final property value ($P_f$), the largest component (LC), number of disconnected components (DC), diameter ($D$), average path length ($\langle l \rangle$), and modularity ($Q$).} 
    \label{tab:final_metrics}

\centering
\footnotesize
\begin{tabular}{c|c|c|c|c|c|c|}
\cline{2-7}
                                            & $P_f$             & LC              & DC          & D            & \textless{}l\textgreater{} & Q               \\ \hline
\multicolumn{1}{|l|}{Clustering}            & 0.416 ± 0.00154  & 4.61e+03 ± 11.6 & 28.8 ± 5.31 & 12.3 ± 0.853 & 4.84 ± 0.0192              & 0.558 ± 0.00435 \\ \hline
\multicolumn{1}{|l|}{Assortative By Degree} & 0.486 ± 0.000574 & 3.93e+03 ± 23.2 & 323 ± 7.65  & 21.6 ± 1.84  & 6.23 ± 0.0526              & 0.572 ± 0.00902 \\ \hline
\multicolumn{1}{|l|}{Assortative By Age}    & 0.974 ± 0.000248 & 4.64e+03 ± 7.3  & 13.8 ± 3.53 & 11.5 ± 0.623 & 4.42 ± 0.0244              & 0.601 ± 0.00782 \\ \hline
\multicolumn{1}{|l|}{Assortative By Race}   & 0.793 ± 0.00271  & 4.65e+03 ± 6.74 & 11.6 ± 3.33 & 8.33 ± 0.53  & 3.54 ± 0.00606             & 0.373 ± 0.00353 \\ \hline
\end{tabular}
\end{table*}

\begin{figure}[h]
\centering
\includegraphics[width=0.7\columnwidth]{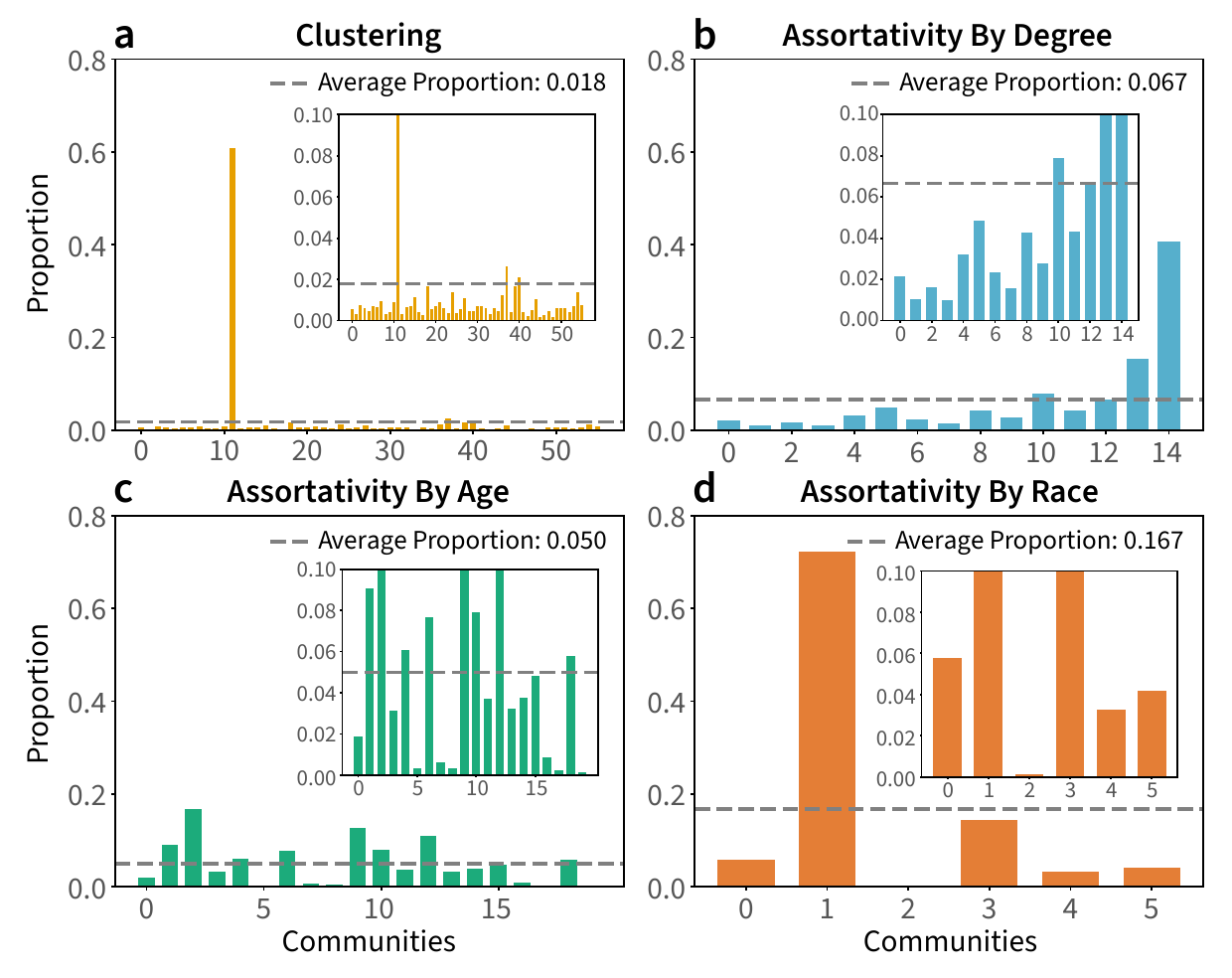}
\caption{\textbf{Number of nodes per community detected by NSBM}. The barplots represent the proportion of nodes in each community for the most clustered (a), and assortative by degree (b), age (c), and race (d) networks. The dashed horizontal line indicates the average node proportion per community.}
\label{fig:node_distribution_community_structure}
\end{figure}

\begin{figure}[h]
\centering
\includegraphics[width=0.7\columnwidth]{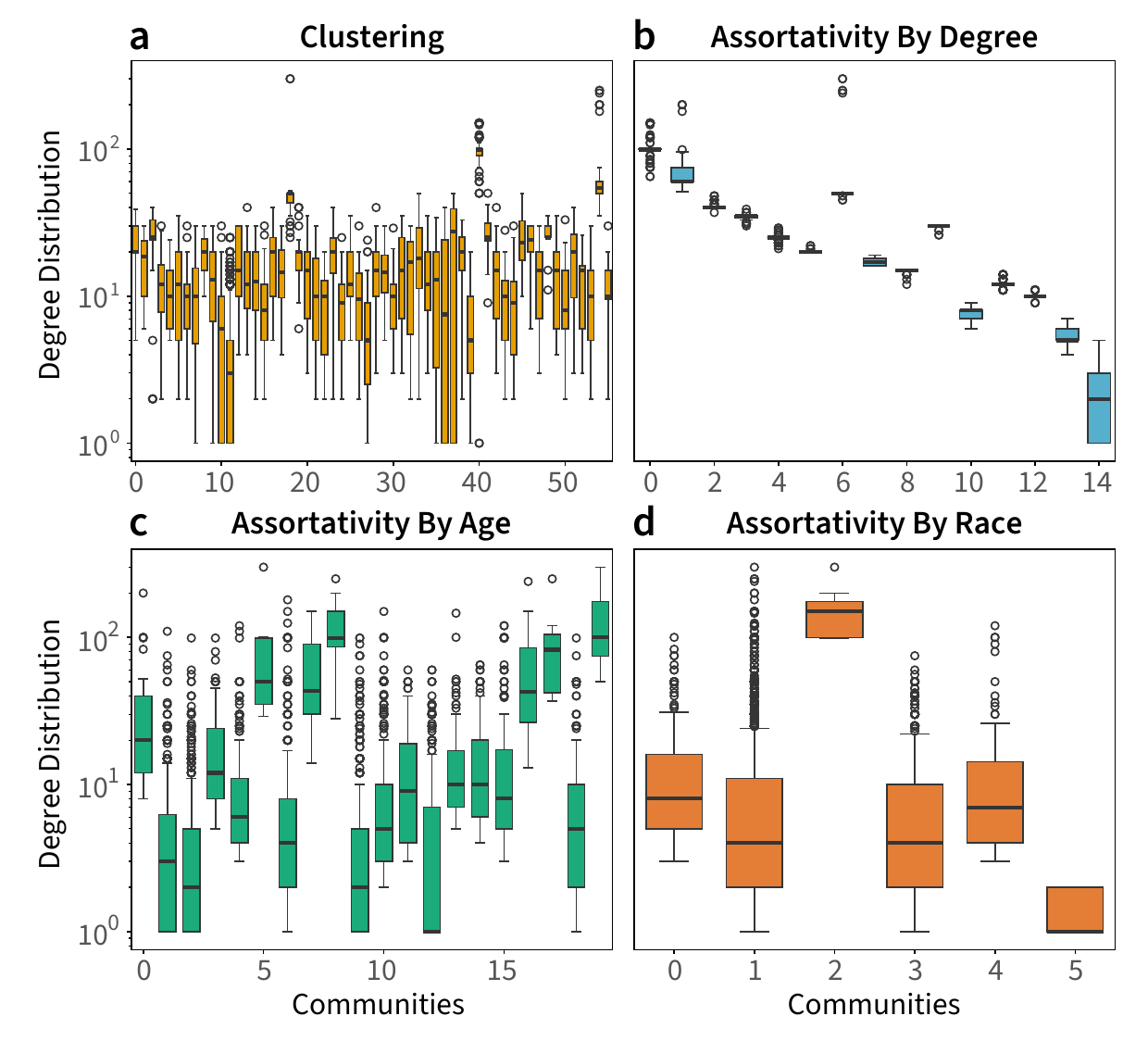}
\caption{\textbf{Node degree distribution within each community detected by NSBM.} Each boxplot presents the median and the interquartile range of the node degrees found within the the most clustered (a), assortative by degree (b), age (c) and race (d) networks. The y-axis is log-scaled for clarity. The outliers detected correspond to nodes whose degree was significantly higher than the median node degree of the respective community (hubs).}
\label{fig:degree_distribution_community_structure}
\end{figure}

At the meso-scale level, we performed community structure analysis using the Nested Stochastic Block Model (NSBM)\cite{peixoto2014hierarchical}. Given the stochastic nature of network generation, we assessed community structure consistency across realizations for each network type using Kolmogorov-Smirnov (KS) tests to compare pairwise community size distributions. Under the null hypothesis that these distributions are identical, most within type comparisons failed to reject $H_0$ ($p>0.05$) (see Supplementary Information, Fig. S7 and S8 and Section 5), indicating strong meso-scale structure similarity among networks generated under the same assortativity and clustering parameter constraints. Therefore, one stochastic trial was chosen to represent the ensemble of networks generated for each network type, from which the community structure was characterized (see Figure \ref{fig:node_distribution_community_structure} and \ref{fig:degree_distribution_community_structure}). While within each network type, there was consistency in the community size distributions, the same distributions revealed great differences when compared across different network types, despite similar modularity values (see Table \ref{tab:final_metrics}).
\par The number of optimal communities varied significantly: clustered networks fragmented into 56 communities, while assortative by race networks formed only 6 (see Figures \ref{fig:node_distribution_community_structure} and  \ref{fig:degree_distribution_community_structure}, and Supplementary Information, Fig. S9). Community size distribution was quite distinct across different networks. In clustered and assortative by degree networks, nodes were concentrated in single large communities (Communities 11 and 14, respectively), while other communities were much smaller than the average community size. Assortative by age networks showed the most homogeneous community size distribution, while assortative by race networks demonstrated the greatest concentration of nodes into few specific communities, with Community 1 containing approximately 70\% of all nodes (see Figure \ref{fig:node_distribution_community_structure}).

We also observed a significantly heterogeneous distribution of highly connected individuals across the generated networks (see Figure \ref{fig:degree_distribution_community_structure}). In assortative by degree and clustered networks, hubs concentrated within communities with few nodes (Community 0 and 1, and Community 18, 41 and 54, respectively), creating isolated transmission chains. In contrast, the largest communities in these networks (Communities 14 and 11, respectively) showed the lowest median degrees. On the other hand, assortative by age and race networks presented hubs distributed more evenly across communities and typically resided in lower-degree communities. This pattern was particularly pronounced in assortative by race networks where despite Community 2 high median degree, most hubs concentrated in the dominant Community 1 (see Figure \ref{fig:node_distribution_community_structure} (d)).

\begin{figure*}[t]
\centering
\includegraphics[width=\textwidth]{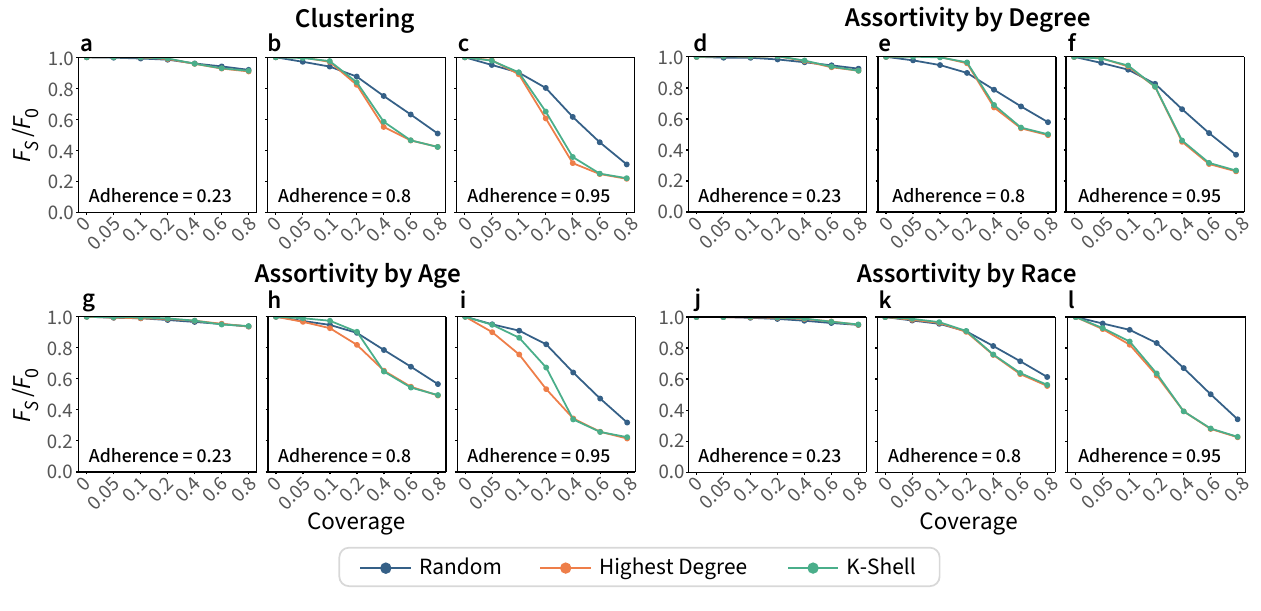}
\caption{\textbf{HIV final size reduction computed for several PrEP distribution scenarios.} HIV final size reduction was defined as the ratio between the final proportion of PLHIV for a certain implemented strategy $F_s$ and the baseline model $F_0$ computed for varying PrEP efficacy (0.23, 0.80, and 0.95), coverage (0.05, 0.10, 0.20, 0.40, 0.60, 0.80), and targeting strategies (random, highest degree centrality, and $k$-shell). Panels a-c, correspond to the most clustered networks, d-f assortative by degree, g-i assortativity by age and j-l assortativity by race.}
\label{fig:HIV_prevalence_reduction_all}
\end{figure*}

\subsection*{Network structure shapes HIV transmission}
The proposed framework to explore extremes of demographic mixing patterns while preserving degree distributions represents a methodological advance that allows us to isolate how different forms of assortativity influence epidemic dynamics. To assess how mesoscopic and global topological properties influence HIV spreading among these sexual contact networks, we simulated HIV transmission dynamics using a discrete-time stochastic agent-based model over 50 years in the largest connected component of each network type (see Methods Section C). Consistent with patterns observed in community structure, the largest component size was stable across network realizations for each type. (see Table \ref{tab:final_metrics} and Supplementary Information, Figure S6 (a)). Accordingly, we used the same network analyzed for community structure to perform our epidemic simulations (see Figure \ref{fig:node_distribution_community_structure} and \ref{fig:degree_distribution_community_structure}). In our model, each node is represented by an individual which is assumed to be in one of the following states: susceptible, infected or on treatment. Transmission probability was highly dependent on the number of infected neighbors, sexual role, condom use (30\% of coverage \cite{kudon2022trends} and 0.70 effectiveness \cite{condom_usage}), and PrEP status. Individuals who are on treatment were assumed virally suppressed and non-infectious \cite{CDC_UequalsU_2024} (see Methods Section C for a full description of the agent-based model implementation). Simulations were initialized with approximately 8.6\% people living with HIV (PLHIV) in the largest component of all networks, with no PrEP intervention at baseline.

Epidemic trajectories were quantified by measuring the proportion of PLHIV, accounting for both infected and on treatment individuals. Baseline simulations revealed substantial variation in equilibrium prevalence:  assortative by degree networks showed the lowest burden (63.2\%, IQR: 62.8--63.7\%) and fastest convergence, while assortative by age and race networks reached the highest prevalence (78.1\%, IQR: 77.5--78.4\% and 80.1\%, IQR: 79.8--80.4, respectively) (see Supplementary Information, Fig. S6 (b)). These differences directly reflect the structural properties characterized above. Assortative by degree networks' increased fragmentation, longer path lengths, and hub isolation created natural transmission barriers, while assortative by race networks' lower modularity facilitated broader spread, a finding with important implications for planning targeted intervention design.

\subsection*{PrEP strategy effectiveness: random versus targeted allocation}

The aforementioned baseline dynamics was used as a benchmark to evaluate PrEP allocation strategies. We systematically varied PrEP individual adherence (0.23, 0.80, 0.95) and coverage (0.05-0.80). The adherence level is proportional to PrEP dosage uptake, such that higher adherence corresponds to lower HIV susceptibility. PrEP coverage corresponds to the fraction of susceptible individuals receiving PrEP. At the beginning of each simulation and for each coverage level, a corresponding fraction of susceptible nodes is selected to be on PrEP. An individual is eligible to be chosen for PrEP uptake if they are HIV-negative \cite{NIH_PrEP}. The selected individuals are assumed to maintain the adoption of the respective prevention strategies during the entire simulation. Additionally, we compared three allocation approaches: random distribution, degree centrality targeting, and k-shell decomposition targeting (see Methods Section E). Strategy effectiveness was quantified as the reduction in cumulative final size of PLHIV relative to baseline ($F_s$/$F_0$) over 50-year simulations (see Figure \ref{fig:HIV_prevalence_reduction_all}).

At low adherence (23\%), all strategies achieved minimal reductions ($F_s/F_0 > 0.90$) regardless of coverage or network type (see Figure \ref{fig:HIV_prevalence_reduction_all} (a), (d), (g) and (j)). At moderate adherence (80\%), targeted strategies demonstrated network-dependent effectiveness. In assortative by age networks, degree-based targeting consistently outperformed random allocation, achieving $F_s/F_0 \approx 0.48$ at 80\% coverage compared to $F_s/F_0 \approx 0.58$ for random distribution (see Figure \ref{fig:HIV_prevalence_reduction_all} (h)). However, clustered and assortative by degree networks exhibited coverage thresholds ($\sim$15\% and $\sim$30\%, respectively) below which random distribution proved superior (see Figure \ref{fig:HIV_prevalence_reduction_all} (b) and (e), respectively). This threshold effect reflects hub distribution patterns: in low coverage settings, targeting hubs in assortative networks by degree and clustered offers less protection to the broader population since only local infections are contained. 

At maximum adherence (95\%), targeted strategies substantially outperformed random allocation across all network types, requiring lower coverage to achieve equivalent reductions in PLHIV. Assortative by age and race networks showed the greatest sensitivity to targeted interventions: degree-based targeting reduced prevalence to $F_s/F_0 \approx 0.30$ at only 40\% coverage in age-assortative networks, while random distribution required 80\% coverage to achieve similar reductions (see Figure \ref{fig:HIV_prevalence_reduction_all} (i) and (l), respectively). Degree centrality consistently outperformed k-shell targeting in assortative by age networks at lower coverage levels ($F_s/F_0 \approx 0.55$ vs. $0.70$ at 20\% coverage), with performance converging at higher coverage (see Figure \ref{fig:HIV_prevalence_reduction_all} (h) and (i)). Such divergence can be explained by the lower correlations between degree and k-shell metrics in assortative by age networks (see Supplementary Information, Fig. S11 (c)).

\begin{figure}[t]
\centering
\includegraphics[width=0.7\columnwidth]{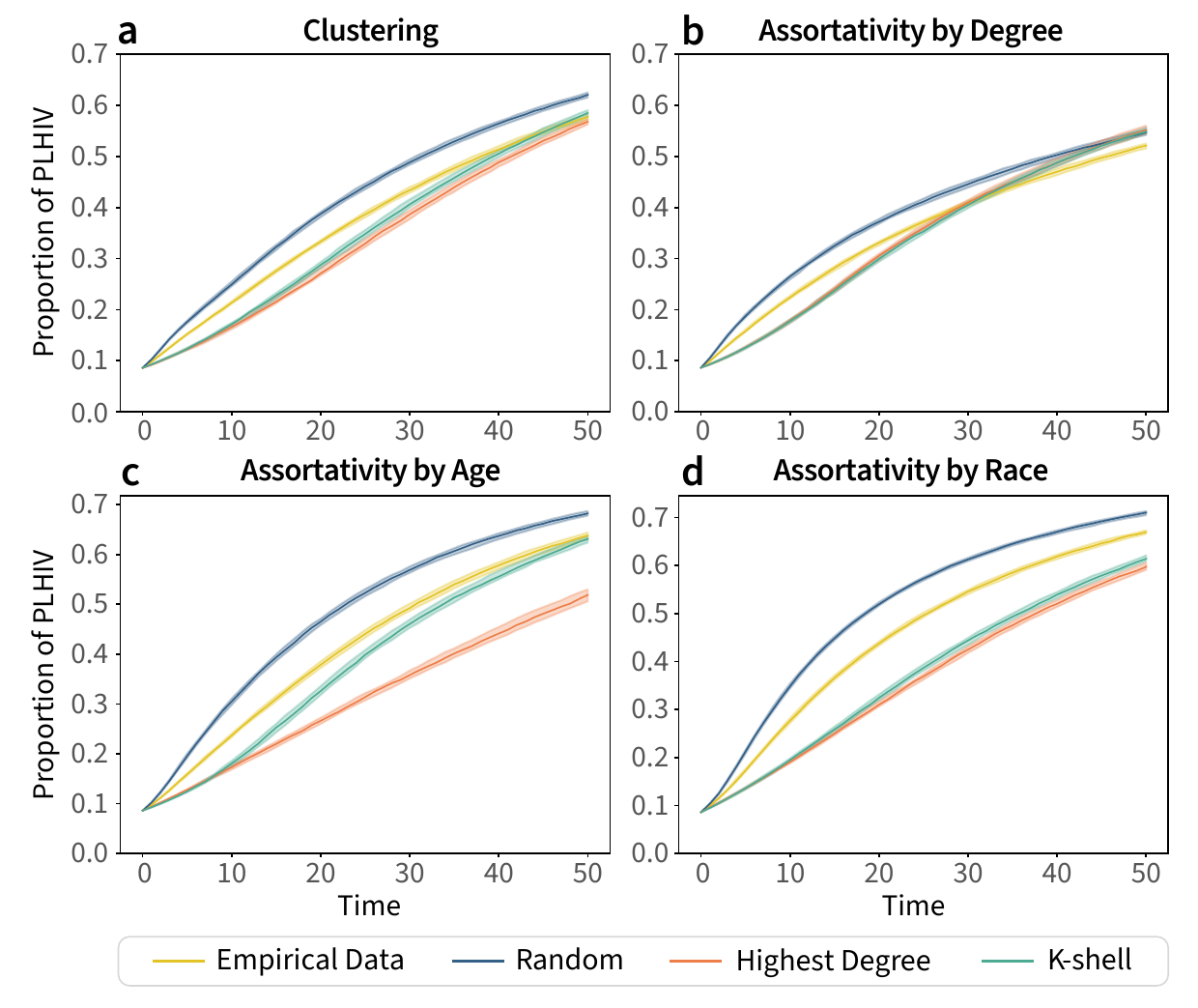}
\caption{\textbf{HIV dynamics comparison between intervention scenarios and current PrEP distribution retrieved from the data.} Proportion of PLHIV across 50 years in clustered (a), and assortative by degree (b), age (c), and race networks (d), considering a PrEP adherence of 95\%. Each colored line corresponds to a different PrEP distribution strategy (empirical PrEP distribution, random, highest degree and k-shell). Prevalence is reported as the median with interquartile range at each time step, based on $N=50$ stochastic simulations.}
\label{fig:HIV_prevalence_all_current_policy}
\end{figure}

Given that PrEP empirical data matches national reported PrEP uptake levels among MSM \cite{CDCPrEPFactSheet}, we also compared PrEP uptake retrieved from the data with the simulated strategies at maximum adherence (95\%), mapping coverage-level observed in the data with those in the simulation (see Figure \ref{fig:HIV_prevalence_all_current_policy}). 
Since network fragmentation varied across topologies, we adjusted coverage to reflect the proportion of PrEP users in each network's largest component. Consequently, in maximally clustered networks, $13.8\%$ of nodes were allocated PrEP, while for networks with maximal assortativity by degree, age and race this percentage was $15.8\%$, $13.8\%$ and $13.7\%$, respectively. Uptake from the data matched or exceeded simulated strategies in clustered and assortative by degree networks, where PrEP coverage exceeded 20\% among medium to high-degree individuals (see Supplementary Information, Fig. S10 (b)). This suggests that current uptake covers network-influential individuals in these topologies. However, in networks where assortativity is driven by age or race, the observed PrEP uptake does not align with network-central positions, suggesting that if real MSM networks exhibit this structure (see Figure~\ref{fig:HIV_prevalence_all_current_policy} (c) and (d)), there may be scope for complementing current broad-access strategies with outreach focused on individuals bridging age or race clusters. In this case, both degree and k-shell targeting substantially outperformed the observed PrEP uptake, particularly in assortative by race networks, where coverage averaged only $\sim$10\% within demographic groups (see Supplementary Information, Fig. S10 (a)). In these networks, hubs bridge demographic communities, yet our results suggest that in these topologies crucial individuals may not be on PrEP. Given that current eligibility criteria are focused on individual risk behaviors, we show that there is an advantage in also identifying these more nuanced transmission-critical individuals. In general, these findings reveal substantial opportunities for optimization through network-informed allocation strategies that explicitly address individuals bridging demographic groups or occupying central network positions, in addition to current guidelines. Moreover, in scenarios where there could be limited resources and PrEP distribution, it is crucial to understand how to optimally distribute it. 


\section*{Discussion}

\par We developed a network reconstruction framework that generates realistic MSM sexual networks from individual-level behavioral and sociodemographic data, typically collected in HIV surveillance studies. While previous approaches have generated networks with enhanced clustering or degree correlations~\cite{alstott2019local,van2010influence,bansal2009exploring}, our study is the first to systematically incorporate and evaluate assortativity by age and race alongside these properties, capturing the demographic homophily that commonly shapes HIV transmission in MSM populations~\cite{abuelezam2019interaction,amirkhanian2014social}. This addresses a critical limitation: most HIV modeling studies either require complete contact data, extremely difficult to obtain for stigmatized populations, or use oversimplified network models that overlook key transmission dynamics.

\par We show that network topology significantly influences both baseline HIV dynamics and intervention effectiveness. Assortative by degree networks showed the lowest equilibrium prevalence and fastest convergence, consistent with prior findings~\cite{chang2020impact,anderson1990significance}. Indeed, the increased fragmentation and longer path lengths in these networks create natural barriers to transmission. However, our study highlights that community structure, not global metrics, emerge as the main determinant of intervention strategies efficiency. In clustered and assortative by degree networks, hubs are concentrated within specific high-degree communities, creating transmission reservoirs where infection circulates rapidly among highly connected individuals~\cite{newman:assortativity_mixing,xulvi2005changing} but spreads slowly to the broader, lower-degree population. Consequently, for moderate adherence levels (80\%), targeted PrEP strategies require higher coverage thresholds to outperform random allocation in these networks. Conversely, in assortative by age and race networks, hubs are distributed across communities and connect diverse demographic groups, positioning them as critical transmission bridges. Here, targeted strategies achieve maximal impact at lower coverage levels. For high adherence levels (95\%) degree and position based allocation reduced prevalence substantially at only 40\% coverage, while random distribution required double this coverage for equivalent effects. These findings are extremely valuable, validating theoretical predictions that identifying influential spreaders maximizes intervention efficiency in heterogeneous networks~\cite{pastor2002immunization,anderson1991infectious}. This can inform public health programs to achieve substantial epidemic control with fewer resources, lower costs, and reduced implementation complexity. 

Among network-based targeting approaches, degree centrality proved more effective than $k$-shell decomposition in assortative by age networks, where hub distribution across communities resulted in fewer discrete $k$-shell values and reduced correlation between metrics, a known limitation of $k$-shell methods in certain network topologies~\cite{da2012predicting,pei2013spreading}. This finding has practical implications: degree data are simpler to collect than $k$-shell indices, which are position-based, yet provide superior targeting performance in demographically assortative networks.

While coverage and optimal PrEP allocation strategies are crucial, adherence is equally critical to intervention success. At low adherence (23\%), all strategies achieved minimal impact regardless of coverage or targeting approach, underscoring that efficient allocation alone cannot compensate for poor adherence. These findings show how crucial health education programs, PrEP counseling, and peer support can be to reduce stigma, discrimination, and misconceptions about PrEP safety and efficacy~\cite{jiang2025preparing}. This could motivate a higher adherence and facilitate network-optimized strategies to achieve their full potential.

\par To assess real-world implementation, we compared PrEP uptake retrieved from the data with simulated strategies, revealing substantial optimization opportunities. When using information about who is on PrEP for network reconstruction, its uptake in the largest component of each network type approximates optimal strategies in assortative-by-degree networks, where coverage exceeds 20\% among high-degree individuals. However, in assortative-by-age and assortative-by-race networks, where hubs bridge demographic groups, both degree and $k$-shell targeting substantially outperformed the empirical distribution, in which lower PrEP coverage is observed within these demographic hubs. Although eligibility criteria typically only consider individual risk behaviors, our results suggest that incorporating network-critical information could significantly improve the identification of individuals whose PrEP initiation and uptake would maximally disrupt transmission chains.To contribute to this, future surveys could go beyond asking about the number of contacts by also collecting data on mixing preferences, which may translate into more effective PrEP allocation strategies. Crucially, because we cannot determine from egocentric data alone which assortativity dimension best reflects the ground-truth network, our findings underscore structural uncertainty as a primary challenge for PrEP efficacy. By clarifying the interplay between network structure and PrEP allocation strategies, this work highlights the importance of investing in richer network data collection to support allocation strategies that remain robust under structural uncertainty.

\par Our framework presents, however, some limitations. Network generation without connectivity constraints resulted in the formation of a giant component and several disconnected components. This fragmentation was most pronounced in assortative by degree networks, where 16\% of nodes were excluded from the largest component. Nevertheless, since real-world sexual networks exhibit similar fragmentation patterns \cite{helleringer2007sexual}, we believe that our modeling approach is still valid. Additionally, imposing connectivity would prevent achieving the high clustering and assortativity coefficients observed empirically~\cite{bansal2009exploring}, reflecting a fundamental trade-off between topological fidelity and completeness.  A key direction for future work is to jointly analyze multiple network properties—such as degree assortativity, clustering, and demographic assortativity by age and race—since in real sexual networks these features are often correlated rather than independent.

Network data remain prohibitively difficult to collect in stigmatized populations. Without real-world contact patterns available, we could not validate our synthetic MSM sexual networks against empirical network structures, limiting assessment of our model's generalizability. Additionally, several HIV surveillance studies present cross-sectional designs \cite{brazia2025assessment}, which limits mathematical modelling of intervention strategies to accommodate dynamic features such as partnership formation and dissolution, shifting concurrency patterns, and potential risk compensation following PrEP adoption. 

\par Overall, our work provides a practical methodology for the study of PrEP distribution optimization when complete network data are unavailable, highly common in HIV surveillance programs. Our findings demonstrate that network-informed allocation strategies could substantially improve HIV prevention efficiency in MSM populations, particularly in networks where hubs bridge demographic groups. These results establish benchmarks for future longitudinal network studies to validate synthetic network approaches and real-world effectiveness. 

\section*{Methods}

\subsection*{A. Dataset Preprocessing}

\indent \par Sociodemographic and behavioral data were preprocessed to enable mathematical modeling of HIV transmission in sexual contact networks. From the initial  4904 MSM that participated in the ARTNet study, we excluded 237 reporting zero sexual contacts, yielding 4,667 individuals. For each participant, we extracted age, HIV status, number of sexual partnerships, race, PrEP status, and testing frequency. For MSM participants with missing PrEP status (82\% of participants), we assumed no PrEP use, resulting in 13.67\% coverage, consistent with 2017 CDC estimates for MSM \cite{CDCPrEPFactSheet}. Missing testing frequency data (18\%) were imputed using k-nearest neighbors (see Supplementary Information Fig. S1 and S2, and Section 1).

\subsection*{B. Network Reconstruction of Null Models to represent Empirical Networks}

\indent \par We consider a graph $G=(V,E)$, in which the elements of $V$ represent the nodes of the network, representing the study participants, connected by edges $e \in E\subseteq V\times V$, representing the sexual interactions between them. The resulting graph (network) obtained from the data preprocessing is composed of $V=4667$ nodes and $E=25107$ edges, such that each node is characterized by the following set of attributes: age, race, number of partnerships, testing frequency and PrEP status. 

\par Since the network structure is unknown, to generate the networks we used the number of partnerships reported by each participant and assume they are connected in the real world. In the context of generating random graphs with a prescribed degree sequence, let $d_i$ represent the degree sequence of graph $G$. Since the sum of all degrees in $G$ denoted by $\sum_{i=0}^{4667} d_i=50214$ is even, random graphs can be constructed pairing up all nodes. We developed an algorithm based on the Havel-Hakimi \cite{havel1955remark,hakimi1962realizability} recursive test that provides a framework to build a simple, undirected, connected
graph meeting the above degree sequence, when the conditions postulated by Erd\H{o}s-Gallai are obeyed \cite{erdos1960graphen,blitzstein2011sequential}. This algorithm was preferred over the classical configuration model since it does not generate self-loops and duplicate edges \cite{barabasi2016network}, which are not desirable to represent sexual networks. In our approach, nodes are ordered in descending order according to their degree, such that higher-degree nodes are the first ones to get paired up. The algorithm stops when the stubs of each node are connected (see Supplementary Information Algorithm 1 and Section 2). 
\par Despite its advantages in the generation of simple graphs, the Havel-Hakimi algorithm is greedy and generates a highly correlated graph configuration space. Therefore, the capability of generalizing the macroscopic structure of our dataset could be compromised. To address this limitation, a MCMC approach was implemented using the method \textit{double edge swap} from NetworkX. Let $G_{t}$ be the graph configuration at time $t$. With probability 0.5, $G_{t+1}$ is determined through an edge swapping process. We select two edges at random $(u,v)$ and $(x,y)$, with distinct endpoints. If $(u,x)$ and $(v,y)$, are not connected, we remove the old edges and replace them by the proposed ones. To maximize exploration of the configuration space, no connectivity constraint was imposed during edge swapping. Since no prior estimates of mixing time exist for such networks, we ran chains until convergence, assessed by evaluating the convergence of the clustering coefficient, degree assortativity, and diameter. The pseudo-code of the algorithm and detailed convergence analysis can be found in Supplementary Information, Fig. S4, Algorithm 2 and Section 3.
\par Null models to represent empirically realistic sexual networks were built using as benchmark the random networks with degree-preserving distributions. We generated networks with enhanced clustering, assortativity by degree, age and race. Degree-preserving algorithms via edge swapping were used such that, the transition probabilities of the Markov chain from states $G_i$ to $G_j$ are
\begin{equation}
    P_{i,j} =
    \begin{cases}
        1 & \text{if property}(G_j) - \text{property}(G_i) > 0 \\
        0 & \text{otherwise},
    \end{cases}
\end{equation}
\noindent in which $G$ corresponds to our network and property to the metric we aim to increase. If the proposed graph configuration $G'$ increased the network property under study, then the Markov chain would transition to a new state. A more detailed description of the algorithms implemented and convergence analysis can be found in Supplementary Information, Fig. S5 and Section 4. 

\subsection*{C. Epidemiological and Agent Based Model} 
\label{sub:contact_network_sctuture}
\indent \par We developed an agent-based model to model HIV disease spreading in the networks with the enhanced clustered and assortative networks by degree, age and race to mimic real-world systems. The model is assumed to be static, which implies that the degree and the nodes each agent interacts with remain constant throughout the entire simulation. The time evolution of the nodes status is modeled by a Markov process. A rejection sampling algorithm was developed to simulate the stochastic process of the disease spreading. Therefore, time was assumed to be discrete and fixed for a $\Delta t=1$ year. Since sexual networks are formed of multiple disconnected components, we did not impose constraints on the network's connectivity in the network generation process. Therefore, the simulation model was applied on the largest component only. 
\par Considering a population of size $N$, each individual is in one of three states: Susceptible $S$, Infected $I$, and Treatment $T$, which corresponds to individuals who start to take Antiretroviral Therapy (ART) after testing positive for HIV. Individuals who are on ART are assumed to reach a zero viral load, in which individuals are not able to transmit the HIV anymore \cite{CDC_UequalsU_2024}. To simulate the HIV transmission we considered $t_{\text{end}}$ = 50 years and $N_{\text{trials}}$ = 50, to reduce bias and assess the robustness of the model to randomness. We chose to implement the stochastic epidemic process in a synchronous way, such that the state of each node is only updated after all nodes have been traversed. Through this approach, each node is given the possibility to transition to another infectious stage. A Bernoulli trial was performed per node to determine if the node will change state. The transition between states will be determined by two main parameters, $\beta$, the probability of a susceptible node to become infected and $\gamma$ the probability of an infected node getting tested and initiating treatment. If the current node is susceptible and presents infected neighbors, such that $u<\beta$, the node transitions from $S \rightarrow I$. There is, however, two types of susceptible individuals in our model, on PrEP and not on PrEP, such that nodes on PrEP, have a lower probability of contracting HIV. Additionally, the infection probability in general also depends on the number of neighbor and the type of interactions, which includes the usage of other prevention methods, such as condoms, and the type of sexual role. The impact of these factors on transition probabilities are discussed in detail below. Finally, if the current node is infected and $u<\gamma$, the node transitions from $I \rightarrow T$. The pseudo-code of this algorithm is provided in the Supplementary Information Algorithm 3.

\subsection*{D. Parameter Estimation}

The testing frequency of each agent retrieved from the data determines the probability of getting tested, expressed as
\begin{equation}
    \gamma=\dfrac {1}{2} \dfrac{\text{Testing Frequency}}{365}.
\end{equation}
\noindent The higher the number of times a person gets tested per year, the higher the likelihood of getting tested. Additionally, since we run simulations with a time increment of $\Delta t=1$ year, we multiply this probability by a 1/2 factor, since each agent reported cumulatively its testing frequency for 24 months.  
\indent \par The computation of the probability of getting infected $\beta$ involves several steps and was based on Vermeer et. al \cite{vermeer2022agent}. First, we compute an overall risk given by:
\begin{equation}
    \text{overall risk}=\beta_0 \; \text{risk}  \times \text{PrEP risk} \times \text{Condom risk},
\end{equation}
\noindent such that the condom and PrEP risk factors are given by subtracting from one the effectiveness of each prevention strategy implemented individually. All individuals are assumed to switch between insertive and receptive roles during the same encounter (versatile). Therefore, the baseline probability of infection per sexual act is given by summing up the probability of contracting the disease by adopting an insertive (0.006) and receptive sexual position (0.014) in the same sexual act \cite{catie}. The probability of HIV transmission through oral sex was assumed to be negligible and therefore considered to have no contribution to the overall transmission risk. Next, we compute the log odds of the overall risk given by:
\begin{align}
\text{log-total-risk} = \ln\left(\frac{\beta_0}{1-\beta_0}\right) + \ln(1 - \text{condom}) \nonumber \\
\quad + \ln(1 - \text{PrEP}).
\end{align}
\noindent Finally, the log-odds is converted again into a probability that will correspond to the overall transmission probability per sexual act given by:
\begin{equation}
    \beta_{\text{per sexual act}}=\dfrac{exp(\text{log-total-risk})}{1+exp(\text{log-total-risk})}.
\end{equation} 
\noindent In order to take into account the heterogeneity of each node's neighbourhood in HIV transmission, a binomial model was defined according to Newman et. al \cite{newman2002spread}. The overall probability of infection $\beta$ depends on the probability that a susceptible node escapes from getting infected, which is constant, and on the number of their infected neighbours $n$. Assuming that time $t$ is discrete we have that:  
\begin{equation}
    \beta=1-(1-\beta_{\text{per sexual act}} \Delta t)^n.
    \label{eq:probability_of_infection}
\end{equation}
\noindent In our epidemiological model, since we incorporate the use of condoms which, contrarily to PrEP usage, contributes to the protection of both nodes involved in a sexual intercourse, some adaptations need to be proposed to equation \ref{eq:probability_of_infection}. Since a susceptible node might not use condoms in all its partnerships, then, the general probability of infection for a given susceptible node is given by:
\begin{equation}
    \beta=1-(1-\beta_{n_+})^{n_+} \times (1-\beta_{n_-})^{n_-}.
\end{equation}
\noindent such that $n_+$ and $n_-$ are the terminologies used to indicate the infected contacts that happen with and without condoms, respectively.

\subsection*{E. Targeted PrEP allocation Strategies}

\par The success of an immunization strategy also depends on the network properties \cite{pastor2002immunization}. Therefore, we compared the effect of choosing nodes uniformly at random with two targeting strategies, based on the node's degree centrality and the k-shell decomposition method. In the degree centrality approach, nodes are prioritized based on their degree, while in the k-shell method, prioritization depends on how nodes are connected and their location in the network. We start by decomposing the network into $k-$shells. First, we remove all the nodes that present only one link, until no more nodes with a remaining degree equal to 1 are present in the network, and we add them to a sub-list which we will denote $1-$shell. Next, we remove all the nodes that have degree 2 and add them to a sub-list called $2-$shell. We continue increasing $k$ and to apply this procedure until all nodes were attributed to a $k-$ shell. The $k-$core corresponds to all $k-$ shells with indices higher or equal to $k$. Nodes that belong to the shell with the highest $k$ form the nucleus \cite{carmi2007model}. These approaches were chosen since, when compared with other centrality measures, they are the most efficient to identify influential nodes in the context of epidemics \cite{de2014role}, especially when the epidemic starts in multiple origins simultaneously \cite{kitsak2010identification}.

\section*{Data Availability}

The dataset used to construct the contact networks is not publicly available. Access is governed by a Memorandum of Understanding (MOU), which specifies the personnel involved and the intended purposes of the data analysis. In accordance with this agreement, the dataset may not be shared without the explicit consent of the ARTnet Study Principal Investigator (Samuel Jenness, Emory University). To request access, please contact the PI via email at \textcolor{blue}{\href{mailto:samuel.m.jenness@emory.edu}{samuel.m.jenness@emory.edu}}. A template MOU will be provided for review, after which access to the ARTnet dataset will be granted through GitHub. Installation and setup instructions are available at \textcolor{blue}{\href{https://github.com/EpiModel/ARTnet}{https://github.com/EpiModel/ARTnet}}.

\section*{Code Availability}

\par The network generation and agent-based model simulations were implemented in Python and can be found at \url{https://github.com/Joao12h/Network-Structure-and-Prevention-Strategies.git}.

\printbibliography

@article{peixoto2014hierarchical,
  title={Hierarchical block structures and high-resolution model selection in large networks},
  author={Peixoto, Tiago P},
  journal={Physical Review X},
  volume={4},
  number={1},
  pages={011047},
  year={2014},
  publisher={APS}
}

@misc{CDCPrEPFactSheet,
  author = {{Centers for Disease Control and Prevention}},
  title = {{PrEP} for {HIV} Prevention in the {US} Factsheet},
  howpublished = {Retrieved from \url{https://www.cdc.gov/nchhstp/newsroom/fact-sheets/hiv/PrEP-for-hiv-prevention-in-the-US-factsheet.html}},
  note = {Accessed on January 8, 2024}
}

@article{de2014role,
  title={Role of centrality for the identification of influential spreaders in complex networks},
  author={De Arruda, Guilherme Ferraz and Barbieri, Andr{\'e} Luiz and Rodriguez, Pablo Martin and Rodrigues, Francisco A and Moreno, Yamir and da Fontoura Costa, Luciano},
  journal={Physical Review E},
  volume={90},
  number={3},
  pages={032812},
  year={2014},
  publisher={APS}
}

@article{catie,
    title = {Putting a Number on It: The Risk from an Exposure to HIV},
    author = {Wilton, James},
    journal = {Canadian AIDS Treatment Information Exchange},
    year = {2012}
}

@article{amirkhanian2014social,
  title={Social networks, sexual networks and HIV risk in men who have sex with men},
  author={Amirkhanian, Yuri A},
  journal={Current HIV/Aids Reports},
  volume={11},
  pages={81--92},
  year={2014},
  publisher={Springer}
}

@article{chang2020impact,
  title={Impact of network assortativity on epidemic and vaccination behaviour},
  author={Chang, Sheryl L and Piraveenan, Mahendra and Prokopenko, Mikhail},
  journal={Chaos, solitons \& fractals},
  volume={140},
  pages={110143},
  year={2020},
  publisher={Elsevier}
}

@article{anderson1990significance,
  title={The significance of sexual partner contact networks for the transmission dynamics of HIV},
  author={Anderson, RM and Gupta, Shivang and Ng, W},
  journal={JAIDS Journal of Acquired Immune Deficiency Syndromes},
  volume={3},
  number={4},
  pages={417--429},
  year={1990},
  publisher={LWW}
}

@misc{hivgov_prep,
  author       = {{HIV.gov}},
  title        = {Pre-Exposure Prophylaxis (PrEP)},
  year         = {2025},
  url          = {https://www.hiv.gov/hiv-basics/hiv-prevention/using-hiv-medication-to-reduce-risk/pre-exposure-prophylaxis},
  note         = {Acessed on March 5, 2025}
}

@article{abuelezam2019interaction,
  title={Interaction patterns of men who have sex with men on a geosocial networking mobile app in seven united states metropolitan areas: observational study},
  author={Abuelezam, Nadia N and Reshef, Yakir A and Novak, David and Grad, Yonatan Hagai and Seage III, George R and Mayer, Kenneth and Lipsitch, Marc},
  journal={Journal of Medical Internet Research},
  volume={21},
  number={9},
  pages={e13766},
  year={2019},
  publisher={JMIR Publications Toronto, Canada}
}

@article{grant2010preexposure,
  title={Preexposure chemoprophylaxis for HIV prevention in men who have sex with men},
  author={Grant, Robert M and Lama, Javier R and Anderson, Peter L and McMahan, Vanessa and Liu, Albert Y and Vargas, Lorena and Goicochea, Pedro and Casap{\'\i}a, Mart{\'\i}n and Guanira-Carranza, Juan Vicente and Ramirez-Cardich, Maria E},
  journal={The New England Journal of Medicine},
  volume={363},
  number={27},
  pages={2587--2599},
  year={2010},
  publisher={Mass Medical Soc}
}

@article{punyacharoensin2016effect,
  title={Effect of pre-exposure prophylaxis and combination HIV prevention for men who have sex with men in the UK: a mathematical modelling study},
  author={Punyacharoensin, Narat and Edmunds, William John and De Angelis, Daniela and Delpech, Valerie and Hart, Graham and Elford, Jonathan and Brown, Alison and Gill, O Noel and White, Richard Guy},
  journal={The Lancet HIV},
  volume={3},
  number={2},
  pages={e94--e104},
  year={2016},
  publisher={Elsevier}
}

@article{vissers2008impact,
  title={The impact of pre-exposure prophylaxis (PrEP) on HIV epidemics in Africa and India: a simulation study},
  author={Vissers, Debby CJ and Voeten, H{\'e}l{\`e}ne ACM and Nagelkerke, Nico JD and Habbema, J Dik F and de Vlas, Sake J},
  journal={PloS One},
  volume={3},
  number={5},
  pages={e2077},
  year={2008},
  publisher={Public Library of Science San Francisco, USA}
}

@article{cdc2021prep,
  title = {{Preexposure Prophylaxis for the Prevention of HIV Infection in the United States - 2021 Update}},
  author = {{US Public Health Service}},
  year = {2021},
  journal = {{Clinical Practice Guideline}}
}

@article{gantenberg2018improving,
  title={Improving the impact of HIV pre-exposure prophylaxis implementation in small urban centers among men who have sex with men: An agent-based modelling study},
  author={Gantenberg, Jason R and King, Maximilian and Montgomery, Madeline C and Gal{\'a}rraga, Omar and Prosperi, Mattia and Chan, Philip A and Marshall, Brandon DL},
  journal={PloS One},
  volume={13},
  number={7},
  pages={e0199915},
  year={2018},
  publisher={Public Library of Science San Francisco, CA USA}
}

@article{kasaie2017impact,
  title={The impact of preexposure prophylaxis among men who have sex with men: an individual-based model},
  author={Kasaie, Parastu and Pennington, Jeff and Shah, Maunank S and Berry, Stephen A and German, Danielle and Flynn, Colin P and Beyrer, Chris and Dowdy, David W},
  journal={JAIDS Journal of Acquired Immune Deficiency Syndromes},
  volume={75},
  number={2},
  pages={175--183},
  year={2017},
  publisher={LWW}
}

@article{pastor2002immunization,
  title={Immunization of complex networks},
  author={Pastor-Satorras, Romualdo and Vespignani, Alessandro},
  journal={Physical Review E},
  volume={65},
  number={3},
  pages={036104},
  year={2002},
  publisher={APS}
}

@article{steinegger2022non,
  title={Non-selective distribution of infectious disease prevention may outperform risk-based targeting},
  author={Steinegger, Benjamin and Iacopini, Iacopo and Teixeira, Andreia Sofia and Bracci, Alberto and Casanova-Ferrer, Pau and Antonioni, Alberto and Valdano, Eugenio},
  journal={Nature Communications},
  volume={13},
  number={1},
  pages={3028},
  year={2022},
  publisher={Nature Publishing Group UK London}
}

@article{kitsak2010identification,
  title={Identification of influential spreaders in complex networks},
  author={Kitsak, Maksim and Gallos, Lazaros K and Havlin, Shlomo and Liljeros, Fredrik and Muchnik, Lev and Stanley, H Eugene and Makse, Hern{\'a}n A},
  journal={Nature Physics},
  volume={6},
  number={11},
  pages={888--893},
  year={2010},
  publisher={Nature Publishing Group UK London}
}

@inproceedings{gkantsidis2003markov,
  title={The Markov chain simulation method for generating connected power law random graphs.},
  author={Gkantsidis, Christos and Mihail, Milena and Zegura, Ellen W},
  booktitle={Alenex},
  pages={16--25},
  year={2003}}

@article{nichols2016cost,
  title={Cost-effectiveness analysis of pre-exposure prophylaxis for HIV-1 prevention in the Netherlands: a mathematical modelling study},
  author={Nichols, Brooke E and Boucher, Charles AB and van der Valk, Marc and Rijnders, Bart JA and van de Vijver, David AMC},
  journal={The Lancet Infectious Diseases},
  volume={16},
  number={12},
  pages={1423--1429},
  year={2016},
  publisher={Elsevier}
}

@article{mayer2020emtricitabine,
  title={Emtricitabine and tenofovir alafenamide vs emtricitabine and tenofovir disoproxil fumarate for HIV pre-exposure prophylaxis (DISCOVER): primary results from a randomised, double-blind, multicentre, active-controlled, phase 3, non-inferiority trial},
  author={Mayer, Kenneth H and Molina, Jean-Michel and Thompson, Melanie A and Anderson, Peter L and Mounzer, Karam C and De Wet, Joss J and DeJesus, Edwin and Jessen, Heiko and Grant, Robert M and Ruane, Peter J and others},
  journal={The Lancet},
  volume={396},
  number={10246},
  pages={239--254},
  year={2020},
  publisher={Elsevier}
}

@article{fonner2016effectiveness,
  title={Effectiveness and safety of oral HIV preexposure prophylaxis for all populations},
  author={Fonner, Virginia A and Dalglish, Sarah L and Kennedy, Caitlin E and Baggaley, Rachel and O’reilly, Kevin R and Koechlin, Florence M and Rodolph, Michelle and Hodges-Mameletzis, Ioannis and Grant, Robert M},
  journal={Aids},
  volume={30},
  number={12},
  pages={1973--1983},
  year={2016},
  publisher={LWW}
}

@article{newman2003mixing,
  title={Mixing patterns in networks},
  author={Newman, Mark EJ},
  journal={Physical Review E},
  volume={67},
  number={2},
  pages={026126},
  year={2003},
  publisher={APS}
}

@article{bansal2009exploring,
  title={Exploring biological network structure with clustered random networks},
  author={Bansal, Shweta and Khandelwal, Shashank and Meyers, Lauren Ancel},
  journal={BMC Bioinformatics},
  volume={10},
  pages={1--15},
  year={2009},
  publisher={Springer}
}

@article{alstott2019local,
  title={Local rewiring algorithms to increase clustering and grow a small world},
  author={Alstott, Jeff and Klymko, Christine and Pyzza, Pamela B and Radcliffe, Mary},
  journal={Journal of Complex Networks},
  volume={7},
  number={4},
  pages={564--584},
  year={2019},
  publisher={Oxford University Press}
}

@book{peixoto2023descriptive,
  title={Descriptive vs. inferential community detection in networks: Pitfalls, myths and half-truths},
  author={Peixoto, Tiago P},
  year={2023},
  publisher={Cambridge University Press}
}

@article{choi2020cost,
  title={Cost-effectiveness analysis of pre-exposure prophylaxis for the prevention of HIV in men who have sex with men in South Korea: a mathematical modelling study},
  author={Choi, Heun and Suh, Jiyeon and Lee, Woonji and Kim, Jun Hyoung and Kim, Jung Ho and Seong, Hye and Ahn, Jin Young and Jeong, Su Jin and Ku, Nam Su and Park, Yoon Soo and others},
  journal={Scientific reports},
  volume={10},
  number={1},
  pages={14609},
  year={2020},
  publisher={Nature Publishing Group UK London}
}

@misc{CDC_UequalsU_2024,
  author       = {{Centers for Disease Control and Prevention}},
  title        = {Undetectable = Untransmittable},
  year         = {2024},
  url          = {https://www.cdc.gov/global-hiv-tb/php/our-approach/undetectable-untransmittable.html},
  note         = {Acessed on December 22, 2024},
  howpublished = {\url{https://www.cdc.gov/global-hiv-tb/php/our-approach/undetectable-untransmittable.html}}
}

@article{peixoto_graph-tool_2014,
         title = {The graph-tool python library},
         url = {http://figshare.com/articles/graph_tool/1164194},
         doi = {10.6084/m9.figshare.1164194},
         urldate = {2014-09-10},
         journal = {figshare},
         author = {Peixoto, Tiago P.},
         year = {2014}}

@book{barabasi2016network,
  title={Network Science},
  author={Albert-László Barabási},
  lccn={2016439537},
  year={2016},
  publisher={Cambridge University Press}
}

@article{xulvi2005changing,
  title={Changing correlations in networks: assortativity and dissortativity},
  author={Xulvi-Brunet, Ram{\'o}n and Sokolov, Igor M},
  journal={Acta Physica Polonica B},
  volume={36},
  number={5},
  pages={1431},
  year={2005}
}

@article{newman:assortativity_mixing,
  title = {Assortative Mixing in Networks},
  author = {Newman, M. E. J.},
  journal = {Phys. Rev. Lett.},
  volume = {89},
  issue = {20},
  pages = {208701},
  numpages = {4},
  year = {2002},
  publisher = {American Physical Society},
  doi = {10.1103/PhysRevLett.89.208701},
  url = {https://link.aps.org/doi/10.1103/PhysRevLett.89.208701}
}

@book{anderson1991infectious,
  title={Infectious diseases of humans: dynamics and control},
  author={Anderson, Roy M and May, Robert M},
  year={1991},
  publisher={Oxford university press}
}

@article{anderson1989mathematical,
  title={Mathematical and statistical studies of the epidemiology of HIV},
  author={Anderson, Roy M},
  journal={Aids},
  volume={3},
  number={6},
  pages={333--346},
  year={1989},
  publisher={LWW}
}

@article{da2012predicting,
  title={Predicting epidemic outbreak from individual features of the spreaders},
  author={Da Silva, Renato Aparecido Pimentel and Viana, Matheus Palhares and da Fontoura Costa, Luciano},
  journal={Journal of Statistical Mechanics: Theory and Experiment},
  volume={2012},
  number={07},
  pages={P07005},
  year={2012},
  publisher={IOP Publishing}
}

@article{weiss2020egocentric,
  title={Egocentric sexual networks of men who have sex with men in the United States: results from the ARTnet study},
  author={Weiss, Kevin M and Goodreau, Steven M and Morris, Martina and Prasad, Pragati and Ramaraju, Ramya and Sanchez, Travis and Jenness, Samuel M},
  journal={Epidemics},
  volume={30},
  pages={100386},
  year={2020},
  publisher={Elsevier}
}

@article{jiang2025preparing,
  title={PrEParing for HIV prevention among men who have sex with men in China: challenges and solutions},
  author={Jiang, Hongbo and Zou, Huachun},
  journal={The Lancet Global Health},
  volume={13},
  number={9},
  pages={e1636--e1641},
  year={2025},
  publisher={Elsevier}
}

@article{erdos1960graphen,
  title={Graphen mit punkten vorgeschriebenen grades},
  author={Erdos, P and Gallai, Tibor},
  journal={Mat. Lapok},
  volume={11},
  pages={264--274},
  year={1960}
}

@article{blitzstein2011sequential,
  title={A sequential importance sampling algorithm for generating random graphs with prescribed degrees},
  author={Blitzstein, Joseph and Diaconis, Persi},
  journal={Internet Mathematics},
  volume={6},
  number={4},
  pages={489--522},
  year={2011},
  publisher={Taylor \& Francis}
}

@article{vermeer2022agent,
  title={Agent-based model projections for reducing HIV infection among MSM: Prevention and care pathways to end the HIV epidemic in Chicago, Illinois},
  author={Vermeer, Wouter and Gurkan, Can and Hjorth, Arthur and Benbow, Nanette and Mustanski, Brian M and Kern, David and Brown, C Hendricks and Wilensky, Uri},
  journal={PloS One},
  volume={17},
  number={10},
  pages={e0274288},
  year={2022},
  publisher={Public Library of Science San Francisco, CA USA}
}

@incollection{world2023hiv,
  title={HIV/AIDS surveillance in Europe 2023--2022 data},
  author={World Health Organization and others},
  booktitle={HIV/AIDS surveillance in Europe 2023--2022 data},
  year={2023}
}

@article{bernini2019evaluating,
  title={Evaluating the impact of PrEP on HIV and gonorrhea on a networked population of female sex workers},
  author={Bernini, Alba and Blouzard, Elodie and Bracci, Alberto and Casanova, Pau and Iacopini, Iacopo and Steinegger, Benjamin and Teixeira, Andreia Sofia and Antonioni, Alberto and Valdano, Eugenio},
  journal={arXiv preprint arXiv:1906.09085},
  year={2019}
}

@article{pei2013spreading,
  title={Spreading dynamics in complex networks},
  author={Pei, Sen and Makse, Hern{\'a}n A},
  journal={Journal of Statistical Mechanics: Theory and Experiment},
  volume={2013},
  number={12},
  pages={P12002},
  year={2013},
  publisher={IOP Publishing}
}

@article{wang2025comparing,
  title={Comparing the Cost-Effectiveness of Alternative Policies for Recommending and Providing HIV Pre-exposure Prophylaxis to Men Who Have Sex With Men in the EU},
  author={Wang, Boxuan and Brazia, Joao and Teixeira, Andreia Sofia and Valdano, Eugenio},
  journal={medRxiv},
  pages={2025--01},
  year={2025},
  publisher={Cold Spring Harbor Laboratory Press}
}

@misc{NIH_PrEP,
  title = {Pre-Exposure Prophylaxis (PrEP)},
  author = {{National Institutes of Health (NIH)}},
  year = {2023},
  url = {https://hivinfo.nih.gov/understanding-hiv/fact-sheets/pre-exposure-prophylaxis-prep},
  note = {Accessed on December 11, 2023}
}

@article{scikit_learn,
  title={Scikit-learn: Machine Learning in {P}ython},
  author={Pedregosa, F. and Varoquaux, G. and Gramfort, A. and Michel, V.
          and Thirion, B. and Grisel, O. and Blondel, M. and Prettenhofer, P.
          and Weiss, R. and Dubourg, V. and Vanderplas, J. and Passos, A. and
          Cournapeau, D. and Brucher, M. and Perrot, M. and Duchesnay, E.},
  journal={Journal of Machine Learning Research},
  volume={12},
  pages={2825--2830},
  year={2011}
}

@article{hakimi1962realizability,
  title={On realizability of a set of integers as degrees of the vertices of a linear graph. I},
  author={Hakimi, S Louis},
  journal={Journal of the Society for Industrial and Applied Mathematics},
  volume={10},
  number={3},
  pages={496--506},
  year={1962},
  publisher={SIAM}
}

@article{havel1955remark,
  title={A remark on the existence of finite graphs},
  author={Havel, V{\'a}clav},
  journal={Casopis Pest. Mat.},
  volume={80},
  pages={477--480},
  year={1955}
}

@article{van2010influence,
  title={Influence of assortativity and degree-preserving rewiring on the spectra of networks},
  author={Van Mieghem, Piet and Wang, Huijuan and Ge, Xin and Tang, Siyu and Kuipers, Fernando A},
  journal={The European Physical Journal B},
  volume={76},
  number={4},
  pages={643--652},
  year={2010},
  publisher={Springer}
}

@article{condom_usage,
  author = {Smith Dawn K. MD MS MPH and Herbst Jeffrey H. PhD and Zhang Xinjiang PhD and Rose Charles E. PhD},
  title = {Condom Effectiveness for {HIV} Prevention by Consistency of Use Among Men Who Have Sex With Men in the United States},
  journal = {JAIDS Journal of Acquired Immune Deficiency Syndromes},
  volume = {68},
  number = {3},
  pages = {337--344},
  year = {2015}
}

@article{newman2002spread,
  title={Spread of epidemic disease on networks},
  author={Newman, Mark EJ},
  journal={Physical Review E},
  volume={66},
  number={1},
  pages={016128},
  year={2002},
  publisher={APS}
}

@article{kudon2022trends,
  title={Trends in condomless sex among MSM who participated in CDC-funded HIV risk-reduction interventions in the United States, 2012-2017},
  author={Kudon, Hui Zhang and Mulatu, Mesfin S and Song, Wei and Heitgerd, Janet and Rao, Shubha},
  journal={Journal of Public Health Management and Practice},
  volume={28},
  number={2},
  pages={170--173},
  year={2022},
  publisher={LWW}
}

@article{carmi2007model,
  title={A model of Internet topology using k-shell decomposition},
  author={Carmi, Shai and Havlin, Shlomo and Kirkpatrick, Scott and Shavitt, Yuval and Shir, Eran},
  journal={Proceedings of the National Academy of Sciences},
  volume={104},
  number={27},
  pages={11150--11154},
  year={2007},
  publisher={National Acad Sciences}
}

@misc{python_louvain,
   title = {python-louvain x.y: Louvain algorithm for community detection},
   author = {Thomas Aynaud},
   year = {2020},
   howpublished = {\url{https://github.com/taynaud/python-louvain}}
 }

@article{helleringer2007sexual,
  title={Sexual network structure and the spread of HIV in Africa: evidence from Likoma Island, Malawi},
  author={Helleringer, Stephane and Kohler, Hans-Peter},
  journal={Aids},
  volume={21},
  number={17},
  pages={2323--2332},
  year={2007},
  publisher={LWW}
}

@article{brazia2025assessment,
  title={Assessment of Knowledge and Use of HIV Primary and Secondary Prevention Strategies in Portugal: A Scoping Review},
  author={Br{\'a}zia, Jo{\~a}o and Wang, Boxuan and Meireles, Paula and Valdano, Eugenio and Teixeira, Andreia Sofia},
  year={2025}
}

\section*{Acknowledgments}

\indent \par J.B. and A. S. T. acknowledge support by FCT – Fundação para a Ciência e Tecnologia – through the LASIGE Research Unit, ref. UID/408/2025. A. P. F. acknowledges also support by FCT through INESC-ID Lisboa, ref. LA/P/0078/2020 (10.54499/LA/P/0078/2020).

\section*{Author Contributions}

A. S. T., A. P. F. and J.B contributed to the conceptualization, methodology, and investigation. J.B. handled data curation, formal analysis, validation, and visualization. A. S. T. and A. P. F. were responsible for funding acquisition, project administration, resources, software, and supervision. I.Z.K. contributed with supervision and to the conceptualization and investigation. All authors discussed and interpreted the results, and wrote the manuscript.




\clearpage
\onecolumn
\appendix

\setcounter{figure}{0}
\renewcommand{\thefigure}{S\arabic{figure}}

\setcounter{table}{0}
\renewcommand{\thetable}{S\arabic{table}}

\setcounter{equation}{0}
\renewcommand{\theequation}{S\arabic{equation}}

\section*{Supplementary Information}

\vspace{0.5cm}

\maketitle
This Supplementary Information provides additional figures and pseudo-code algorithms that support and clarify the results presented in the main manuscript. It also includes detailed descriptions of the data imputation techniques, the network reconstruction algorithms and their convergence analyses, as well as the rationale behind the selection of the partition methods used to characterize the community structure of the generated networks.

\vspace{1cm}
\tableofcontents
\clearpage

\renewcommand{\thesection}{\arabic{section}}
\setcounter{section}{0}

\section{Section 1: KNN Imputation}

Data imputation was performed using the \textit{KNNImputer} from the \textit{Scikit-learn} library \cite{scikit_learn}, which fills in missing values based on the mean of the $K$ nearest neighbors in the dataset. In order to be handled by the algorithm, we performed one hot encoding of the categorical variables and standardization and normalization of  the numerical attributes. Additionally, we determined the correlation matrix of all selected variables (see Figure S1). We verified that there is a relevant correlation between the testing frequency and being currently on PrEP, which is expected since according to the U.S. PrEP guidelines, PrEP acess requires frequent HIV testing every 3 months \cite{hivgov_prep}. Since the proportion of missing values was higher among non-PrEP users compared to PrEP users, we split the participants into two cohorts based on PrEP status and applied imputation separately to avoid bias and the potential overestimation of each participant testing frequency. We tested several values of $K$ and found that $K=5$ preserved most of the empirical distribution before and after KNN imputation (see Figure S2).

\section{Section 2: Havel-Hakimi}

\par The proposed Havel-Hakimi algorithm receives as an input a list $D$ of tuples, such that each tuple corresponds to the ID of the current node and its respective degree. Per iteration, $D$ is ordered in descending order such that the tuple that presents the highest degree is removed from $D$. This node will be denoted by ego. When the ego is removed, the resultant list $D'$ will be composed by the remaining nodes and respective degrees. After, the selected node will be connected sequentially to the other nodes, such that the nodes that present a higher degree have priority. This explains why the Havel-Hakimi algorithm presents itself as a greedy algorithm. As the current ego is connected to each node from $D'$, the number of available stubs is decreased such that their degrees are updated, respectively, to $D[0][1]-1$ and $D'[i][1]-1$, where $i$ corresponds to the current node from $D'$ selected to be connected with the ego. When all stubs from the current ego are connected ($D[0][1]=0$), the current ego is updated and becomes the first node from $D'$. For the degree sequence to be graphical, the degrees of successive $D'$ sequences obtained by removing the ego from $D$ should always be superior or equal to zero. Otherwise, the number of edges of the degree sequences changes for the algorithm to produce a graphical sequence, which explains why, in the final step, we verify if the number of edges of the generated graph with the prescribed degree sequence $L$ remains equal to the number of initial edges. We generated 10 networks through the Havel-Hakimi algorithm.

\section{Section 3: MCMC Convergence Analysis}

\noindent \par
\begin{figure}[H]
\centering
\includegraphics[width=1.0\textwidth]{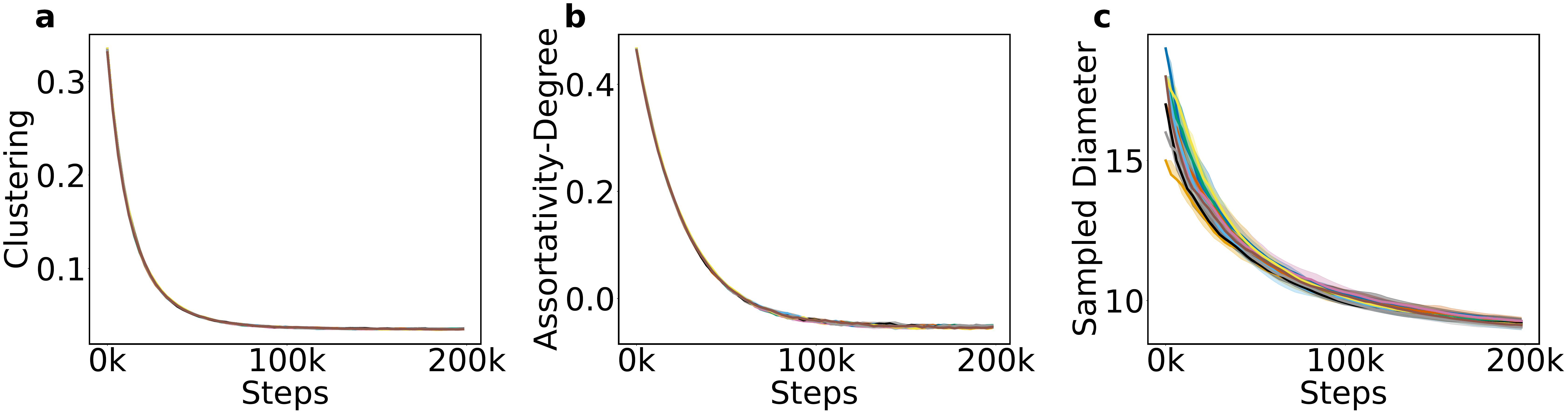}
\caption{\textbf{Markov Chain convergence obtained for ten different networks generated through the Havel-Hakimi algorithm.} We generated 10 MCMC networks for each Havel-Hakimi network during 200,000 steps. The median and respective interquartile range of successful edge swappings were 91030.0 (90883.25-91181.5). The figure presents the median and interquartile variation of the transitivity (a), assortativity by degree coefficient (b) and the sampled diameter (c) throughout the simulation.}
\label{fig:markov_chains}
\end{figure}

 To assess the convergence of the uniform sampling following a MCMC approach, we measured the transitivity, assortativity by degree, and the sampled diameter \cite{gkantsidis2003markov} (see Figure \ref{fig:markov_chains}) of 100 random networks, sampled from 10 initial networks obtained from the Havel-hakimi algorithm. Initially, due to the greedy nature of the Havel-Hakimi, networks present high transitivity, assortativity by degree, and diameters. However, as the number of successful edge swaps increased, the networks reached a stationary state characterized by vanishing clustering and assortativity coefficients, as well as reduced diameters, indicating a transition towards random mixing.

\par Finally, similarly to the markov chain approach, we generated clustered, assortative by degree, age and race networks through an edge swapping process. The evolution of the clustering and assortativity coefficients by attribute as the number of successful edge swapping increases is displayed in Figure S4. Most networks have converged after 200,000 steps, such that clustered and assortative networks by degree reached faster a plateau phase. In addition, networks that were produced by increasing the assortativity by age present an almost perfect assortative mixing.

\section{Section 4: Clustered and Assortative Networks Reconstruction}

\indent \par Similarly to the generation of uniform and independent networks, to reproduce clustered and assortative  by degree, age and race networks, we performed edge rewiring to preserve the network's degree. 
\par Several approaches were developed to increase clustering and assortativity by attribute. To increase the clustering of the network, we proposed an approach to maximize the number of triangles proposed per iteration and also the number of successful edge swappings to reduce the time required to reach the network's convergence similar to what was proposed by Alstott et. al \cite{alstott2019local}. First, two random nodes $u$ and $x$ are selected. Next, we obtain all the combinations of node pairs between a node $v$ and $y$ such that $v \in N(u)$ and $y \in N(x)$ and we compute the common neighborhood $N(v,y)$ for all edge pairs. The probability an edge pair will be chosen to be proposed as a new edge will be proportional to the number of common neighbors it presents since the higher the number of common neighbors, the higher the formation of triangles per iteration, and, consequently, the global clustering increases. After verifying if the proposed edges $(u,x)$ and $(v,y)$ induce the formation of self-loops or multi-edges, the number of common neighbors of the selected edges to be switched $N(u,x)$ and $N(v,y)$ are computed. If $N(u,x)+N(v,y)>N(u,v)+N(x,y)$, then the new graph configuration is accepted and the edges are swapped. To maximize the assortativity by degree coefficient of the network, we followed the approach proposed by Van Mieghem et al \cite{van2010influence}, such that if $(d_u - d_x)^2+(d_v - d_y)^2 < (d_u - d_v)^2+(d_x - d_y)^2$, we accept the proposed edges. For the assortativity by age and race, the methods \textit{numeric\_assortativity\_coefficient} and \textit{numeric\_assortativity\_coefficient} and \textit{attribute\_assortativity\_coefficient} from \textit{NetworkX} were used, which were based in the formulas deduced by Newman \cite{newman2003mixing} to compute assortativity for numerical and categorical attributes, respectively. 

\section{Section 5: Community Detection}

\indent \par Community detection is frequently used to provide insights regarding the structure of complex and large networks. One of the most popular algorithms to characterize community partitions is the Louvain algorithm, a fast modularity maximization algorithm. However, modularity maximization algorithms present some limitations. Particularly, they have a resolution limit, and consequently, find at most $\sqrt{2E}$ groups in connected networks, which means that merges the smaller communities together, but also split the larger community into several spurious ones. Therefore underfitting and overfitting of different parts of the network might happen simultaneously \cite{peixoto2023descriptive}. Furthermore, since it's a descriptive method, the network generative process is not taken into account and, consequently, the detected communities might be completely random. 
\par
\begin{figure}[H]
\centering
\includegraphics[width=1.0\textwidth]{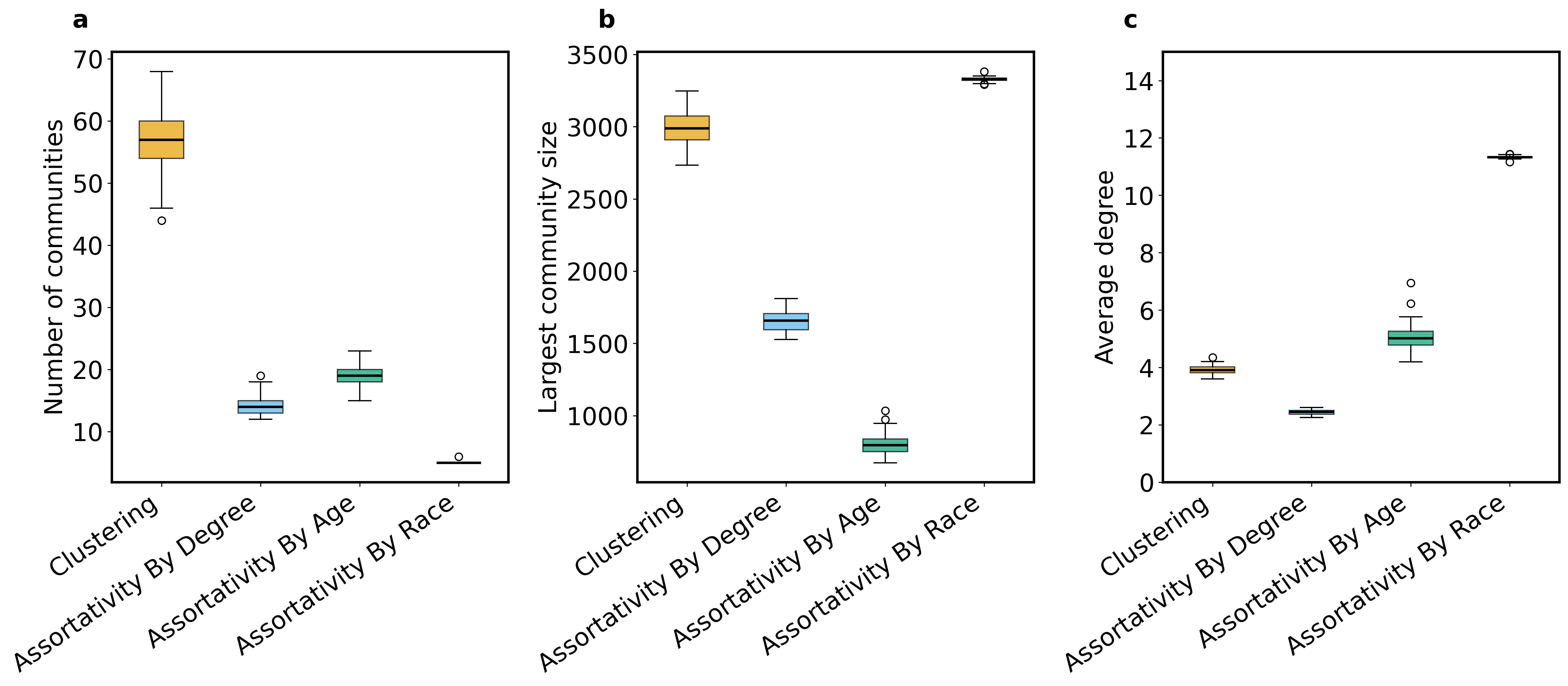}
\caption{\textbf{Structural characterization of community partition obtained by the Nested Stochastic Block Model over 100 networks realizations for each network type.} We present, for each network type, the median and interquartile range of the number of communities (a), largest community size (b) and average degree (c). }
\label{fig:community_statistics}
\end{figure}

\begin{figure}[H]
\centering
\includegraphics[width=1.0\textwidth]{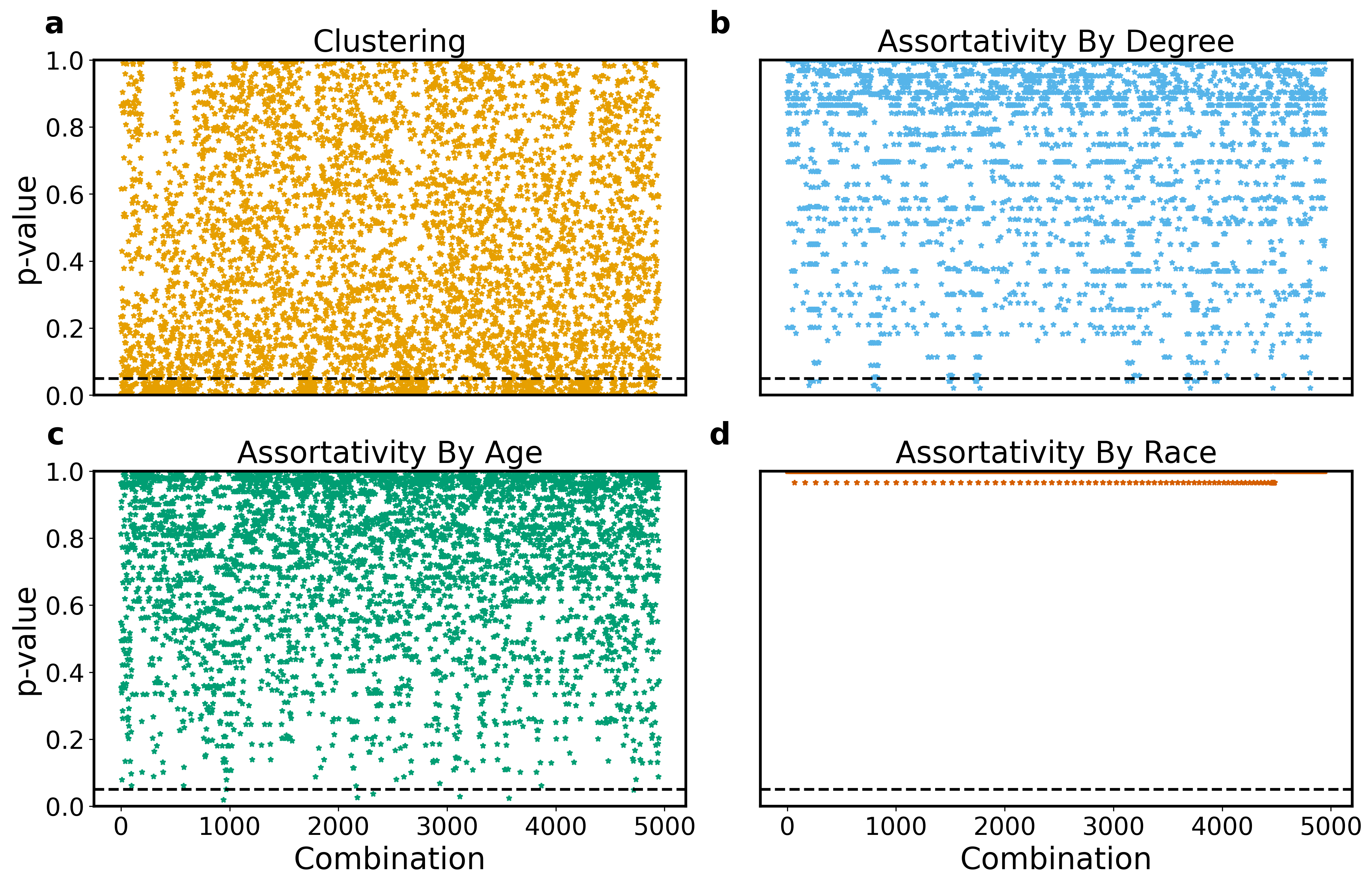}
\caption{\textbf{Two-Sample Kolmogorov–Smirnov Test on the community size distribution for the partitions obtained for the 100 realizations of each network type.} We present the p-values obtained for each pairwise combination of each distribution of number of nodes across partitions for clustered (a), assortativity by degree (b), age (c) and race (d) networks. The dashline represents the threshold below which the null hypothesis $H_0$ that any two distributions come from the same distribution should be rejected.}
\label{fig:pvalues_community_size}
\end{figure}

 We applied the Nested Stochastic Block Model (NSBM), a probabilistic approach to address the reported limitations of modularity maximization algorithms. For each stochastic realization of the 100 clustered, assortative by degree, age and race networks, we obtained $N=50$ partitions using the NSBM. For each stochastic trial and network type, we chose the partition that registered the lowest descriptive length \cite{peixoto_graph-tool_2014}. We evaluated the robustness of the partitions obtained for each network stochastic trial by obtaining the distribution of the number of communities detected, size of the largest community and average degree (see Figure \ref{fig:community_statistics}), which was very similar across network types for different partitions obtained for each stochastic trial. To quantify this consistency, we applied two-sample Kolmogorov--Smirnov tests to community size distributions from 100 realizations per network type (Figure \ref{fig:pvalues_community_size}). Under the null hypothesis that pairwise community size distributions come from the same distribution, most comparisons within each network type failed to reject $H_0$ ($p>0.05$): 83.9\% for clustered networks, 99.4\% for degree-assortative, 99.88\% for age-assortative, and 100\% for race-assortative networks.

\section{Figures}

\begin{figure}[H]
\centering
\includegraphics[width=0.60\textwidth]{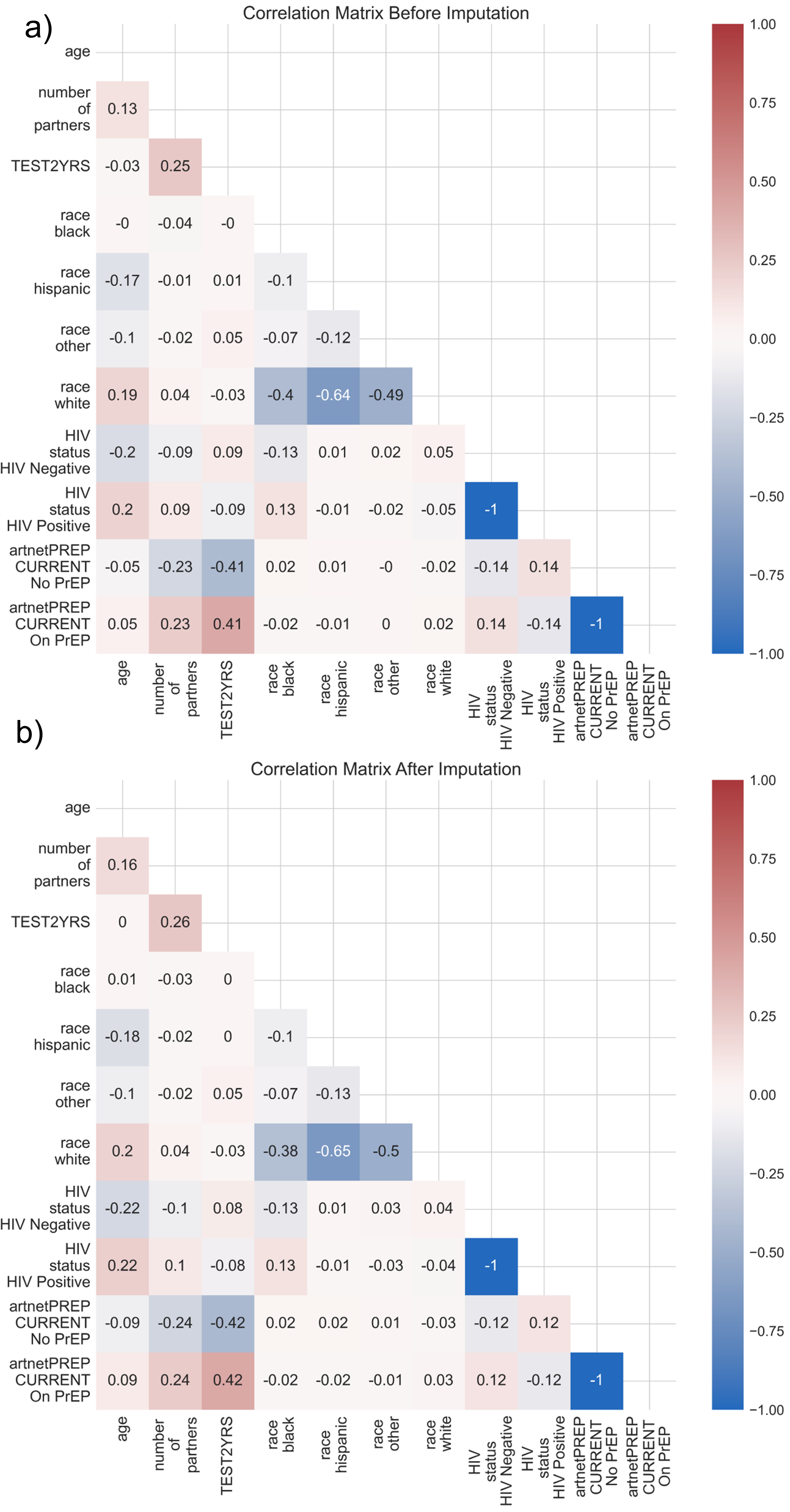}
\caption{\textbf{Feature Correlation between dataset variables.} Correlation matrix  comparison before (a) and after (b) applying KNN imputation $(K=5)$ to handle missing data.}
\label{fig:correlation_matrix}
\end{figure}

\begin{figure}[H]
\centering
\includegraphics[width=0.7\textwidth]{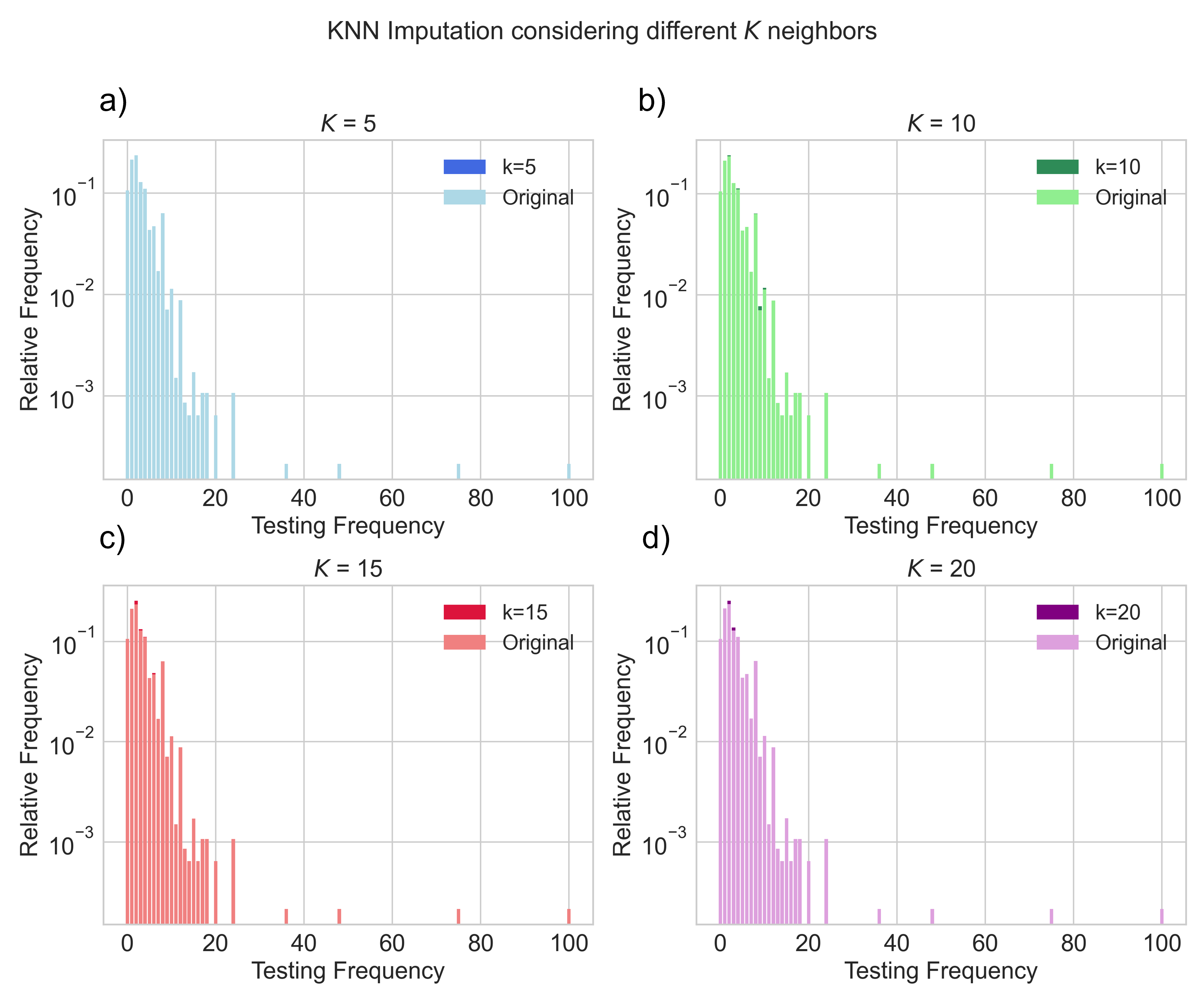}
\caption{\textbf{Testing Frequency Distribution distribution before and after data imputation.} Imputation was performed for (a) $K=5$, (b) $K=10$, (c) $K=15$ and (d) $K=20$. The y-scale is shown on a logarithmic scale to improve data visualization.}
\label{fig:KK_imputation}
\end{figure}

\begin{figure}[H]
\centering
\includegraphics[width=0.7\textwidth]{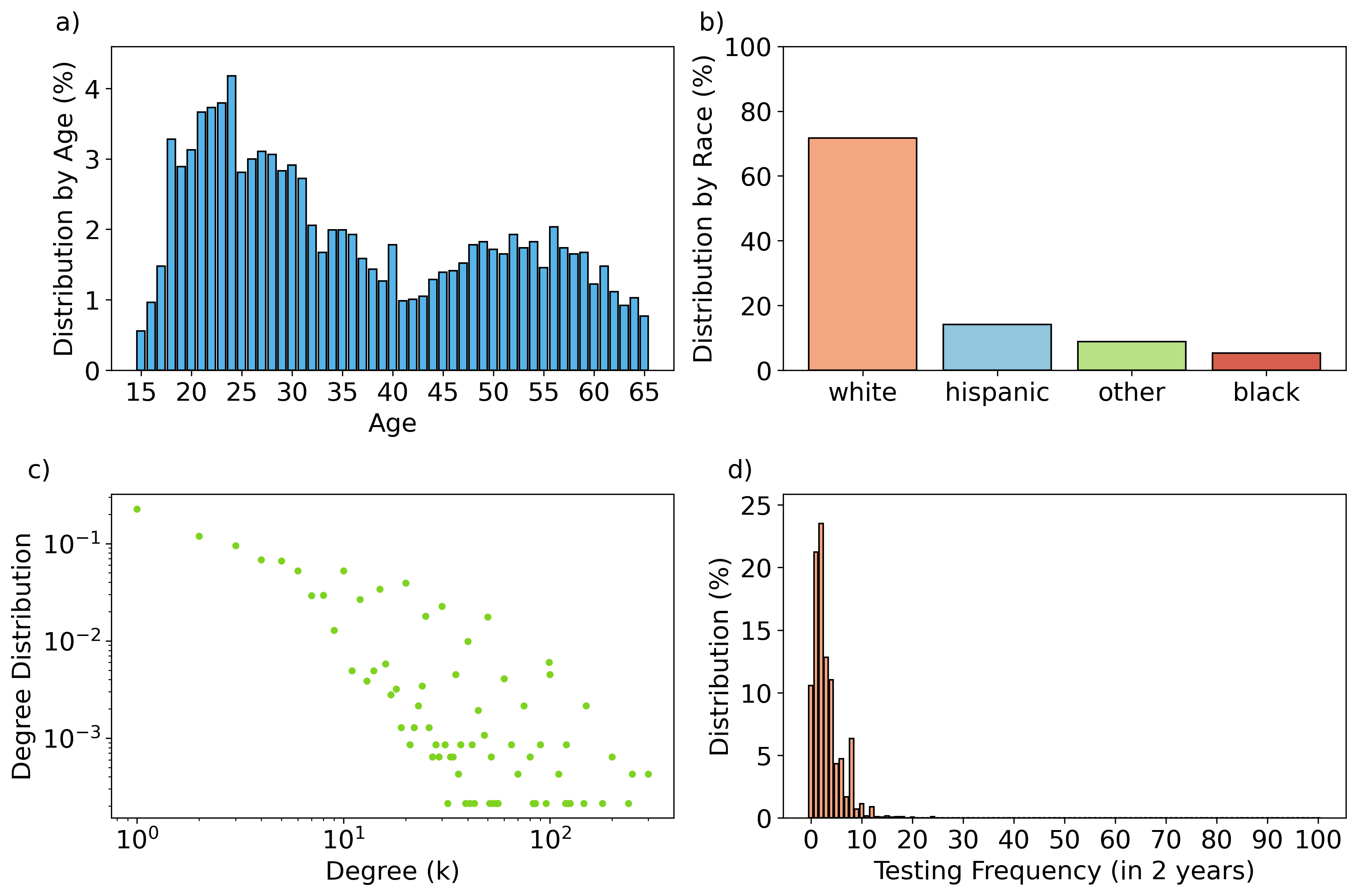}
\caption{\textbf{Dataset Statistical Characterization.} Panel (a) corresponds to the node distribution by age, (b) to node distribution by race, (c) to degree distribution and (d) to node testing distribution. The degree distribution was highly skewed, therefore, panel c is presented in a log-scale to improve visualization and interpretability.}
\label{fig:dataset statistics}
\end{figure}

\begin{figure}[H]
\centering
\includegraphics[width=0.7\textwidth]{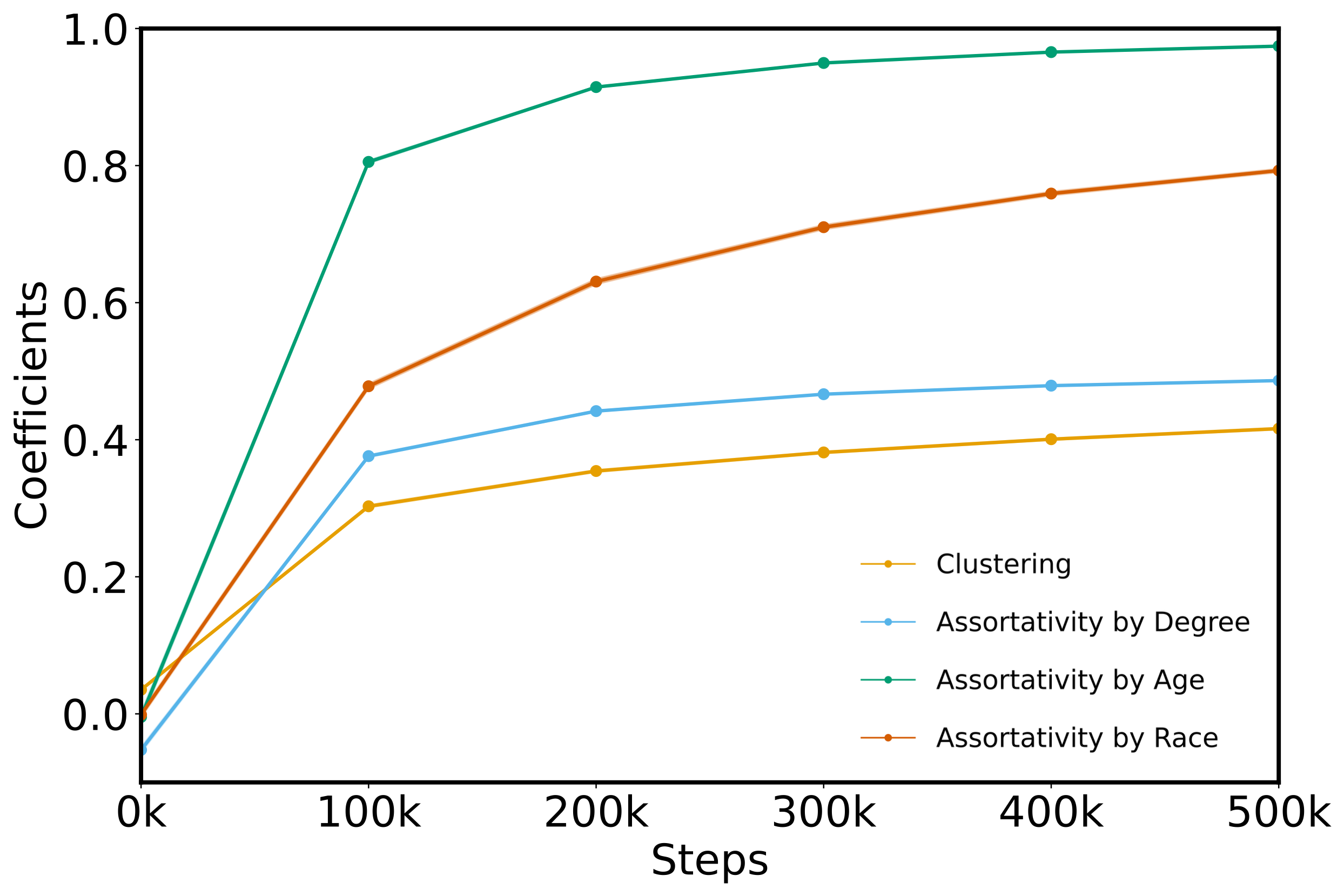}
\caption{\textbf{Network properties coefficient evolution during network reconstruction based on clustered and assortativity by degree and sociodemographic attributes.} The figure presents the median and interquartile variation of 100 clustered, assortativity by degree, age and race networks, simulated for 500,000 steps.} 
\label{fig:structured_networks_evolution_per_edge_swappings}
\end{figure}

\begin{figure}[H]
\centering
    \begin{subfigure}[t]{0.95\textwidth}
        \centering
        \includegraphics[width=\textwidth]{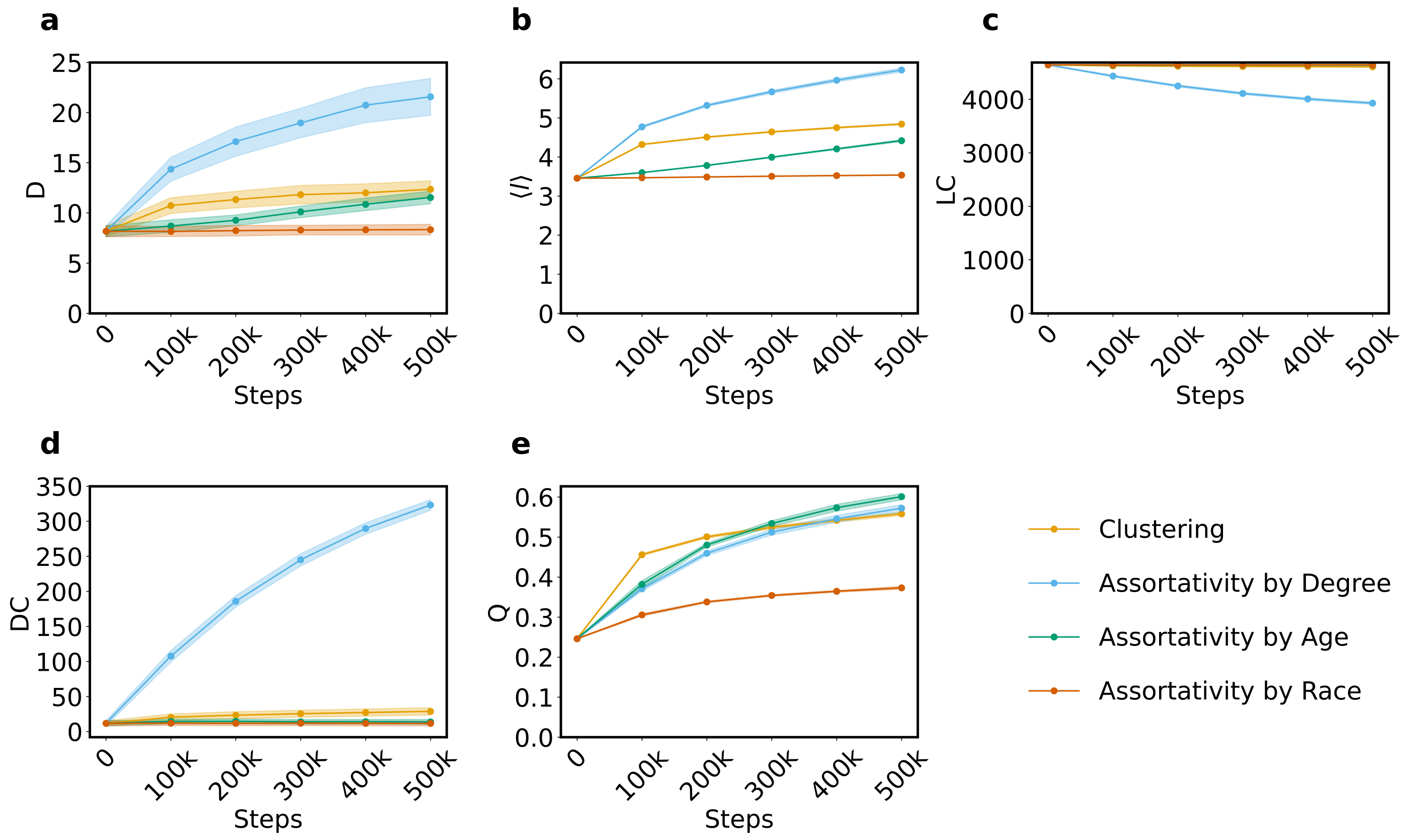}
        \caption{Evolution of structural metrics (diameter, average shortest path length, largest component size, number of disconnected components, and modularity) as clustering and assortativity coefficients increase over 500,000 steps. Each global property was represented through their respective mean and standard deviation over 100 MCMC networks obtained for each network type.}
        \label{fig:structure_evolution}
    \end{subfigure}
    
    \vspace{0.5em}
    
    \begin{subfigure}[t]{0.50\textwidth}
        \centering
        \includegraphics[width=\textwidth]{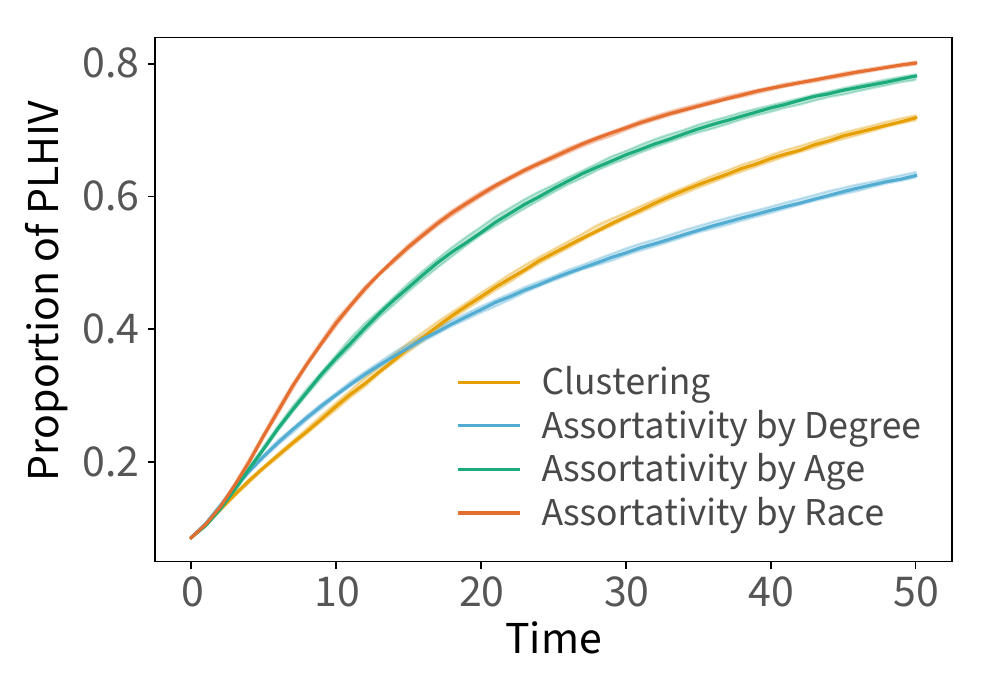}
        \caption{HIV prevalence over 50 years across networks with different topologies. Each line shows the median and interquartile range from $N = 50$ stochastic simulations.}
        \label{fig:HIV_prevalence}
    \end{subfigure}

\caption{\textbf{Relationship between network structure and epidemic outcomes.} (a) Structural network properties under increasing clustering and assortativity. (b) HIV prevalence dynamics in clustered and assortative networks.}
\label{fig:combined_vertical}
\end{figure}

\begin{figure}[H]
\centering
\includegraphics[width=0.7\textwidth]{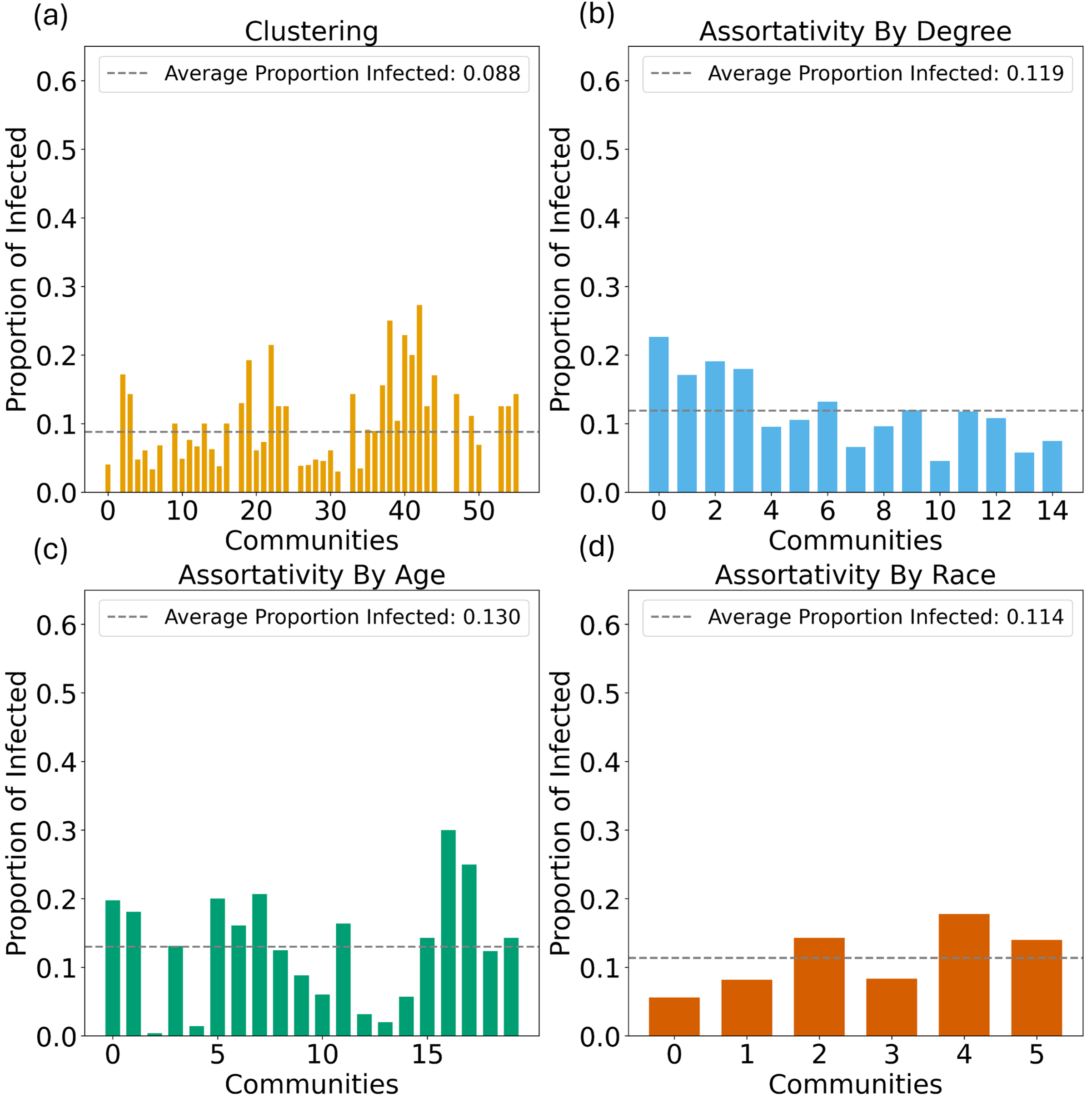}
\caption{\textbf{HIV prevalence per community detected by the NSBM.} Proportion of infected nodes adjusted to the community size detected by the NSBM for the most clustered (a), assortative by degree (b), age (c), and race (d) networks. The dashed horizontal line corresponds to the average proportion of infected per community for each network.}
\label{fig:infected_nodes_proportion}
\end{figure}

\begin{figure}[H]
\centering
\includegraphics[width=1.0\textwidth]{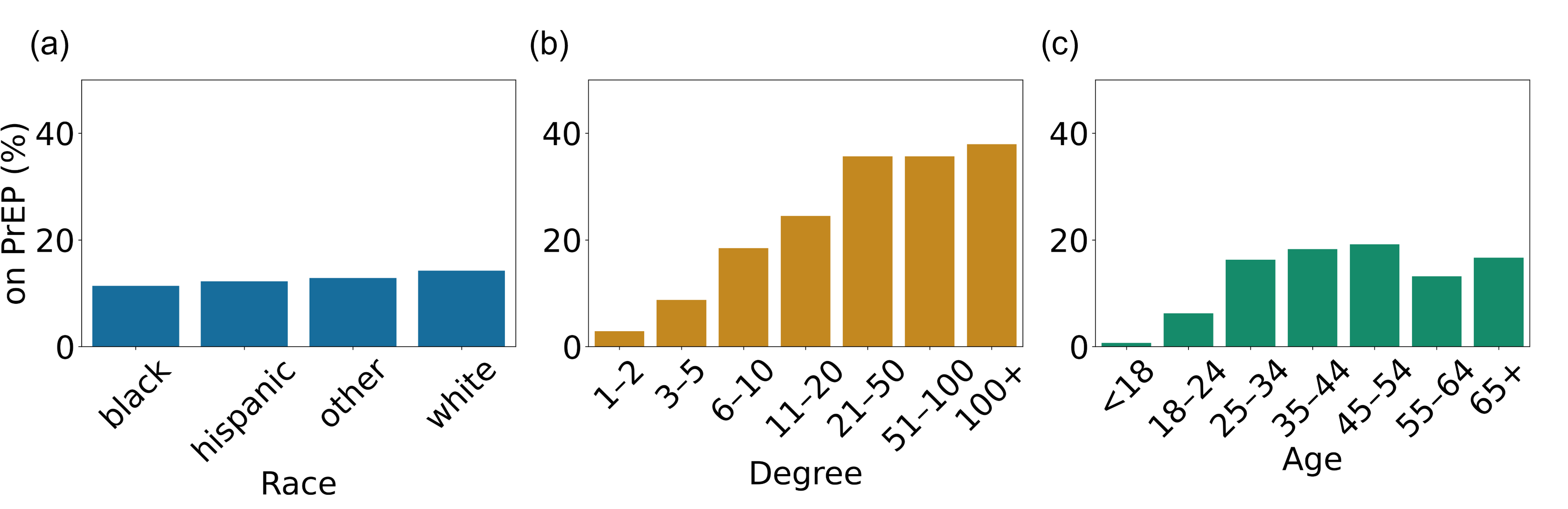}
\caption{\textbf{PrEP Distribution among dataset participants.} We characterize the proportion of PrEP users retrieved from empirical data according to race (a), degree (b) and age (c) in the largest component of the most assortative networks by race, degree and age, respectively. The proportion of PrEP users was adjusted to each group size.}
\label{fig:prep_distribution}
\end{figure}

\begin{figure}[H]
\centering
\includegraphics[width=0.7\textwidth]{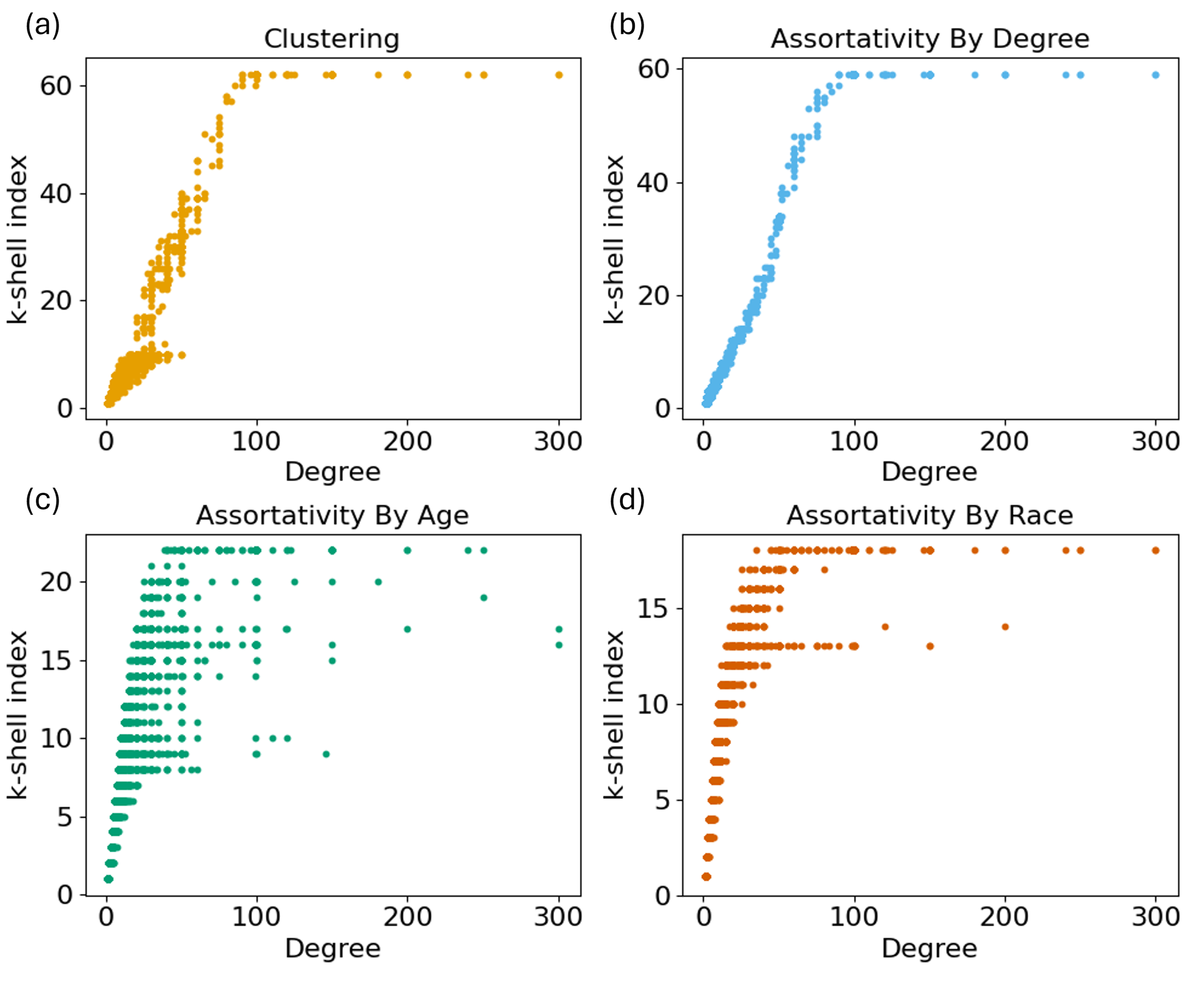}
\caption{\textbf{K-Shell and Degree Centrality Correlation}. We assessed the correlation between the number of k-shells detected by the k-shell decomposition algorithm and the degree of each node for the most clustered (a), assortative by degree (b), age (c), and race (d) networks.}
\label{fig:k-shell correlation}
\end{figure}

\section{Algorithms}

\begin{algorithm}[H]
\caption{Havel Hakimi Algorithm}
\label{alg:havel_hakimi}
\begin{algorithmic}
\State \textbf{Input:} List $D$ of tuples.
\State \textbf{Output:} List $L$, $len(L)==\# edges$.
\State Initialize empty list L;
\While{$D$ is not empty}
    \State $random.shuffle(D)$
    \State Order $D$ from highest to lowest degree;
    \State Select and Remove $D[0]$;
    \State index=0;
    \State Initialize empty list $S$;
    \While{$D[0][1]>0$}
        \State Append $(D[0][1],D[index][1])$ to $S$;
        \State $D[0][1]=D[0][1]-1$;
        \State $D[index][1]=D[index][1]-1$;
        \State i=i+1;
    \State Append $S$ to $L$.
    \EndWhile
\EndWhile
\end{algorithmic}
\end{algorithm}

\begin{algorithm}[H]
\caption{Markov Chain Monte Carlo Sampling}
\label{alg:cap}
\begin{algorithmic}
\State \textbf{Input:} graph $G$, $N$ iterations.
\State \textbf{Output:} graph $G$, List of transitivities $C$, List of Assortativity Coefficients by Degree Coefficients $A_d$, List of Sampled Diameters $d$,Successful edge swaps $swapcount$.
\State $C$=[current transitivity];
\State $d$=[current diameter];
\State $A_d$=[current assortativity coefficient by degree];
\State swapcount=0;
\State iteration=1;
\While{$iteration<N$}
    \State Generate random number $u \in [0,1]$;
    \If{$u<0.5$:}
        \State Select pair of nodes $(u,x)$ based on cumulative degree distribution;
        \If{$u!=x$:}
            \State Select uniformly at random vertex $v$ from $N(u)$;
            \State Select uniformly at random vertex $y$ from $N(x)$;
        \EndIf
        \If{$v!=y$}
            \If{$x$ not in $N(u)$ and $y$ not in $N(v)$}
                \State Remove edge pairs $(u,v)$ and $(x,y)$ and replace them by edge pairs $(u,x)$ and $(v,y)$;
                \State $swapcount$+=1;
            \EndIf
        \EndIf
    \EndIf
    \State Update current transitivity, assortativity by degree coefficient, and sampled diameter;
    \State iteration+=1.
\EndWhile
\end{algorithmic}
\end{algorithm}

\begin{algorithm}[H]
\caption{Stochastic Simulation HIV Spreading}
\label{alg:synchronous_update}
\begin{algorithmic}

\State \textbf{Input:} Current graph $G$, maximum time $t_{end}$, number of trials $N_{trials}$, $\Delta t$.
\State \textbf{Output:} List of lists [times, $S$, $I$, and $T$] such that each entry corresponds to a stochastic trial and gives the number in each state at each time.
\State Compute the Largest Component of the current graph $G$;
\State $\text{trial}=0$;
\State Get all nodes of the largest component;
\While{$\text{trial}<N_{trials}$}
    \State $times,S,I,T = [0],[S_0],[I_0],[T_0]$;
    \State $\text{current time}=\Delta t$;
    \State Compute list of HIV status, sexual roles, and PrEP adherence for  all nodes;
    \State Compute list of infected neighbors for each susceptible node;
    \While{$\text{current time}<t_{end}$}
        \State Add current time to times;
        \State Create a new list of HIV status and infected neighbors for each susceptible to be updated;
        \For{node in all nodes}
            \State Generate random number $u \in [0,1]$;
            \If{node is infected and $u<\gamma$}
                \State Remove node from infected neighbors of a given susceptible;
                \State Update HIV status of the current node to Treatment;
                \State $I_0,T_0 = I_0-1,T_0+1$;
            \ElsIf{node is susceptible and presents infected neighbours}
                \State Compute probability of remaining susceptible after contact with $n$ infected neighbors;
                \If{$u<1-\text{probability of remaining susceptible}$}
                    \State Add node to infected neighbors list of a given susceptible;
                    \State Update HIV status of the current node to Infected;
                    \State $S_0,I_0=S_0-1,I_0+1$;
                \EndIf
            \EndIf
        \EndFor
        \State Update current HIV state vector to updated respective list;
        \State Update current $I$ neighbors per $S$ to updated respective list;
        \State current time+=$\Delta t$;
        \State Update $S$, $I$, and $T$;
    \EndWhile
    \State trial+=1.
\EndWhile
\end{algorithmic}
\end{algorithm}

\end{document}